\documentclass[acmlarge, screen]{acmart}

\AtBeginDocument{%
  \providecommand\BibTeX{{%
    \normalfont B\kern-0.5em{\scshape i\kern-0.25em b}\kern-0.8em\TeX}}}

\setcopyright{rightsretained}
\acmJournal{IMWUT}
\acmYear{2024} \acmVolume{8} \acmNumber{2} \acmArticle{78} \acmMonth{6}\acmDOI{10.1145/3659623}

\usepackage{graphicx}
\usepackage{subcaption}
\usepackage{amsmath}
\usepackage{pifont}

\DeclareMathOperator*{\argmax}{arg\,max}

\usepackage{xspace}
\usepackage[utf8]{inputenc}
\usepackage{enumitem}
\newcommand{\name}{G-VOILA\xspace}
\newcommand{\system}{VOILA-G\xspace}
\usepackage{xcolor}
\usepackage{mdframed}

\newcommand{\RII}[1]{\textcolor{black}{#1}}
\newcommand{\RIII}[1]{\textcolor{black}{#1}}
\newcommand{\RIV}[1]{\textcolor{black}{#1}}
\newcommand{\Concise}[1]{\textcolor{black}{#1}}

\newcommand{\takeaways}[1]{\vspace{0.0cm}\begin{mdframed}[backgroundcolor=gray!15]#1\end{mdframed}\vspace{0.4cm}}
\begin{document}

%\title{\name: Understanding User Behavior with Gaze-Facilitated Information Retrieval in Daily Scenarios}

\newpage

\title[\name: Gaze-Facilitated Information Querying in Daily Scenarios]{\name: Gaze-Facilitated Information Querying in Daily Scenarios}

% \name: gaze-driven information querying in daily scenarios 

\author{Zeyu Wang}
\email{wang-zy23@mails.tsinghua.edu.cn}
\orcid{0009-0007-5048-1665}
\affiliation{%
  \institution{Key Laboratory of Pervasive Computing, Ministry of Education, Department of Computer Science
and Technology, Tsinghua University}
  \streetaddress{30 Shuangqing Rd}
  \city{Haidian Qu}
  \state{Beijing Shi}
  \country{China}
}

\author{Yuanchun Shi}
\authornote{Corresponding authors.}
\orcid{0000-0003-2273-6927}
\affiliation{%
  \institution{Key Laboratory of Pervasive Computing, Ministry of Education, Department of Computer Science
and Technology, Tsinghua University}
  \streetaddress{30 Shuangqing Rd}
  \city{Haidian Qu}
  \state{Beijing Shi}
  \country{China}
}
\affiliation{%
  \institution{Intelligent Computing and Application Laboratory of Qinghai Province, Qinghai University}
  \streetaddress{}
  \city{Xining}
  \state{Qinghai}
  \country{China}
}
\email{shiyc@tsinghua.edu.cn}

\author{Yuntao Wang}
\authornotemark[1] 
\orcid{0000-0002-4249-8893}
\affiliation{%
  \institution{Key Laboratory of Pervasive Computing, Ministry of Education, Department of Computer Science
and Technology, Tsinghua University}
  \streetaddress{30 Shuangqing Rd}
  \city{Haidian Qu}
  \state{Beijing Shi}
  \country{China}
}
\email{yuntaowang@tsinghua.edu.cn}

\author{Yuchen Yao}
\orcid{0009-0000-7954-7372}
\affiliation{%
  \institution{Tsinghua University}
  \streetaddress{30 Shuangqing Rd}
  \city{Haidian Qu}
  \state{Beijing Shi}
  \country{China}
}
\email{yaoyc19@mails.tsinghua.edu.cn}

\author{Kun Yan}
\orcid{0000-0001-8290-5169}
\affiliation{%
  \institution{Microsoft Research Asia}
  \streetaddress{}
  \city{Haidian Qu}
  \state{Beijing Shi}
  \country{China}
}
\affiliation{%
  \institution{SKLSDE Lab, Beihang University}
  \streetaddress{}
  \city{Haidian Qu}
  \state{Beijing Shi}
  \country{China}
}
\email{kunyan@buaa.edu.cn}

\author{Yuhan Wang}
\orcid{0009-0003-5139-1267}
\affiliation{%
  \institution{Beijing University of Posts and Telecommunications}
  \streetaddress{}
  \city{Haidian Qu}
  \state{Beijing Shi}
  \country{China}
}
\email{yigetianluoturanjiu@gmail.com}

\author{Lei Ji}
\orcid{0000-0002-7569-3265}
\affiliation{%
  \institution{Microsoft Research Asia}
  \streetaddress{}
  \city{Haidian Qu}
  \state{Beijing Shi}
  \country{China}
}
\email{leiji@microsoft.com}

\author{Xuhai Xu}
\orcid{0000-0001-5930-3899}
\affiliation{%
  \institution{Massachusetts Institute of Technology}
  \streetaddress{}
  \city{Cambridge}
  \state{Massachusetts}
  \country{United States}
}
\email{orson.xuhai.xu@gmail.com}

\author{Chun Yu}
\orcid{0000-0003-2591-7993}
\affiliation{%
  \institution{Tsinghua University}
  \streetaddress{30 Shuangqing Rd}
  \city{Haidian Qu}
  \state{Beijing Shi}
  \country{China}
}
\email{chunyu@tsinghua.edu.cn}

% \renewcommand{\shortauthors}{Wang, et al.}

\begin{abstract}
Modern information querying systems are progressively incorporating multimodal inputs like vision and audio. However, the integration of gaze --- a modality deeply linked to user intent and increasingly accessible via gaze-tracking wearables --- remains underexplored.
This paper introduces a novel gaze-facilitated information querying paradigm, named \name, which synergizes users’ gaze, visual field, and voice-based natural language queries to facilitate a more intuitive querying process. 
In a user-enactment study involving 21 participants in 3 daily scenarios (p = 21, scene = 3), we revealed the ambiguity in users' query language and a gaze-voice coordination pattern in users' natural query behaviors with \name. 
Based on the quantitative and qualitative findings, we developed a design framework for the \name paradigm, which effectively integrates the gaze data with the in-situ querying context.
Then we implemented a \name proof-of-concept using cutting-edge deep learning techniques. A follow-up user study (p = 16, scene = 2) demonstrates its effectiveness \RIII{by achieving both higher objective score and subjective score,} compared to a baseline without gaze data. 
We further conducted interviews and provided insights for future gaze-facilitated information querying systems.

\end{abstract}



% \received{15 Nov 2023}

\maketitle

\section{introduction}
\label{sec:intro}

As the cyber and physical spaces quickly merge, people exhibit a significant demand for information retrieval (IR) anywhere and anytime in their daily lives~\cite{buschel2018demonstrating, here_and_now, modern-information-retrieval, dick2021immersive-learning, jitir}, no longer confined to a specific device or location. 
With advancements in the computational capabilities of wearable devices, the incorporation of a virtual assistant that can provide on-demand, in-situ answers to users' inquiries has the potential to greatly facilitate the interaction with surrounding targets ~\cite{wang2022faceori, yan2023conespeech} and enhance the naturalness of the user's information retrieval experience~\cite{ajanki2010contextual}. Specifically, smart glasses with gaze tracking open new possibilities for natural information retrieval techniques in daily scenarios by combining the voice and gaze modalities~\cite{land2006eye-movements-and-actions, tanriverdi2000interacting-with-eye}.

Prior research has investigated the potential of utilizing smart glasses ~\cite{gao2023mmtsa, wang2023modeling} equipped with gaze tracking to streamline information retrieval related to physical objects~\cite{here_and_now, ajanki2010contextual, Piening_2021looking-for-info, content_placement, lin2021labeling, xair, yan2023voilaa}. 
These approaches enhanced information retrieval by superimposing digital widgets around physical entities, allowing users to access information through interacting with the widgets in a multi-stage manner~\cite{buschel2018demonstrating, ajanki2010contextual}. While this has been a significant stride, the ultimate goal is to enable users to retrieve information using natural expressions, eliminating the need for these intermediary widgets. 

This paper introduces \name\footnote{\RIII{We refer to our proposed paradigm as \name because it can be an abbreviation of \textbf{G}aze-facilitated \textbf{VOI}ce and \textbf{VI}sual-based natural \textbf{L}anguage querying \textbf{A}ssistant. Also, ``Voila'' is used to indicate or draw attention to something that is being presented or brought to someone's notice in French, meaning ``there it is'' in English, which can represent the case of using pronouns as described in Section~\ref{sec:language_pattern}}}, a future information querying paradigm that combines users' gaze data, visual field and voice-based natural language queries. However, a challenge arises when users employ multi-modal, often ambiguous expressions during interactions with virtual assistants tailored for information querying in everyday settings~\cite{put-that-there, liu2023understanding-in-situ}. The central issue lies in the precise interpretation of such ambiguous expressions, particularly in the presence of data from other modalities, such as gaze. Previous work have shown the relation between eye movements pattern and daily activities~\cite{bulling2009eye, land2006eye-movements-and-actions}, there remains a gap in understanding how gaze is intricately linked to daily querying, especially in the context of mouth-eye coordination.

\begin{figure}
    \centering
  \includegraphics[width=\textwidth]{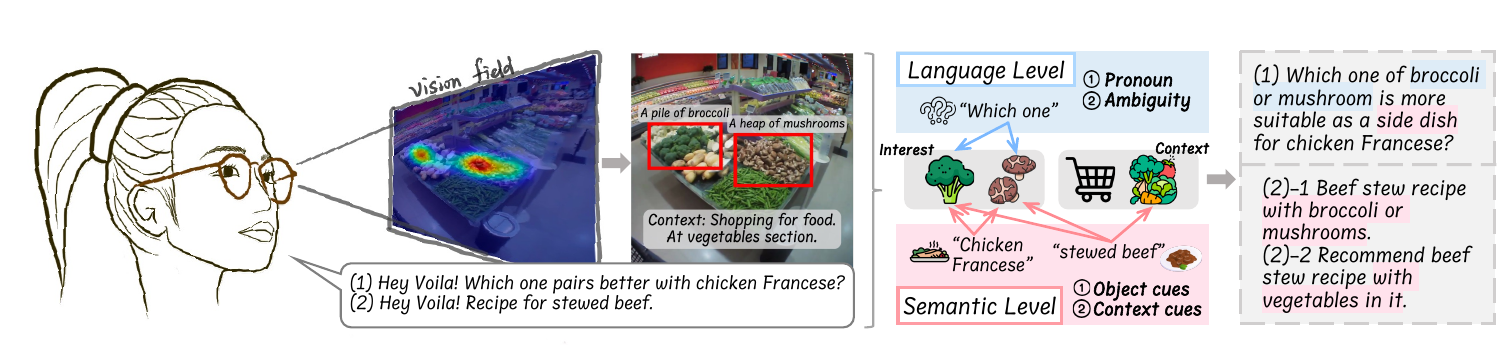}
  \caption{Illustrating \name use cases in a shopping scenario. A user wearing smart glasses is shopping for dinner. She reaches the vegetable aisle and desires detailed information about specific vegetables. Here is how an assistant within the \name{} paradigm might assists her: (1) \textbf{Query Initiation}: She posts queries through natural voice commands. (2) \textbf{Contextual Analysis}: The assistant analyzes the user's field of view and gaze to discern specific areas of interest. (3) \textbf{Query Response}: By further aligning the situational contexts with the posted questions, the assistant deduces her precise query intent and delivers a clear response.}
  \Description{}
  \label{fig:teaser}
\end{figure}

To gain insights into users' natural expression patterns, anticipated use cases, and potential engagement with \name{}, we conducted a user-enactment study involving 21 participants.
Our findings indicate that users frequently omitted specific context-related details in their expressions, assuming that the system could inherently understand such contexts. 
The study also unveiled distinct patterns of mouth-eye coordination in user expressions, particularly arising from pronoun usage and gaze fixation on items relevant to their queries. Figure~\ref{fig:teaser} illustrates how the user naturally interacts with \name paradigm in a supermarket for ingredient selection.

Based on the identified natural expression patterns from the user-enactment study, we proposed a design framework for implementing an intelligent assistant within the \name paradigm. Specifically, we established an inferential pipeline between the user's gaze pattern and the often ambiguous intent behind their inquiries.

To evaluate the efficacy of the \name paradigm and proposed framework, we implemented a preliminary version of \RIII{\system\footnote{\RIII{We put the letter ``G'' backward to separate the \name paradigm from our preliminary implementation, also to be in consist with other baselines mentioned in Section~\ref{sec:baselines}}}} assistant to conduct a proof-of-concept user study, with 16 participants using Pupil Labs Invisible gaze tracker~\cite{tonsen2020high} in 2 daily life settings, covering 12 specific inquiry tasks. We compared \system to baseline methods in an ablation manner. The results indicated that the answers generated by \system achieved an 89\% objective recall score, surpassing the baseline by 11\%, illustrating the increment of incorporating gaze in an information querying system. \system also received a significantly higher level of preference score than the baseline in terms of subjective evaluation matrices, including matchness between reconstructed query expression and user intent, as well as satisfaction, precision, usefulness for the generated response, etc. We gained insights in an afterwards interview session to further understand user's experience and anticipation for \system. 

Overall, our contributions can be summarized as four-fold:
\begin{itemize}
    \item We proposed a future information querying paradigm that combines users' gaze data, visual field and voice-based natural language question as input, namely the \name paradigm.

    \item We conducted an user-enactment study that reveals users' natural query behavior in a mouth-eye coordination manner within the \name paradigm, as well as revealed qualitative and quantitative findings.

    \item We proposed a design framework for intelligent assistant within the \name paradigm upon insights gained from the user-enactment study. We implemented a proof-of-concept assistant under the proposed framework, namely \system.

    \item We conducted a user study that validated \system's effectiveness in inferring users' query intent and showed significantly better subjective outcomes across several metrics including satisfaction, usefulness, and mental demand etc. % the effectiveness and showed a user preference for information querying with \name.

\end{itemize}
\section{Related Works} 
\label{sec:related_works}
In this section, we describe the related work, including information retrieval for daily scenarios, contextual computing for interaction, and multi-modal large models. We discuss the difference between \name and existing works at the end of each subsections. 

\subsection{Information Retrieval for Daily Scenarios}
Information Retrieval (IR) has evolved from manual document searches to keyword-based online search engines, yet still encounters query reformulation issue~\cite{modern-information-retrieval, query-reformulation}. Recently, generative AI has empowered search engines to understand more natural and expressive queries~\cite{microsoft_bing}. Meanwhile, multi-modality-based search tools have started to leverage diverse information types as input~\cite{google-multisearch}. Despite these advancements, contemporary IR techniques remain predominantly restricted to traditional screen-based interfaces.

With the advancement of Augmented Reality (AR) devices, researchers have integrated IR closely with physical world through Head-Mounted Displays (HMDs)~\cite{ajanki2010contextual, buschel2018demonstrating, Piening_2021looking-for-info, xair, lin2021labeling, spatiality-and-semantics, ir-tourist}, which is defined as reality-based information retrieval (RBIR) by ~\cite{here_and_now}. RBIR adheres to the principle of ``find what you need with zero query terms or less'' ~\cite{allan2012frontiers}, adopting a proactive strategy to impose information on certain objects in order to reduce inquiry costs~\cite{lin2021labeling, spatiality-and-semantics, ir-tourist}. However, this approach introduces the issue of information overload and challenges in precisely addressing users' queries.

\Concise{To tackle information overload in AR, Piening et al~\cite{Piening_2021looking-for-info} proposed gaze-adaptive interfaces to modulate the display of information panels. Meanwhile, researchers ~\cite{here_and_now, buschel2018demonstrating, ajanki2010contextual} introduced multi-round IR operations leveraging AR's interactivity to facilitate precise information acquisition, including selection and search refinement. For example, a user interested in avocados at a grocery store may select ``nutrition details'' rather than ``recipes'' on avocados' information panel. However, these proactive display strategies either struggle to align with user intent or require extensive interaction, impeding the immediate recognition of information utility\cite{jitir}. In contrast, natural language queries can convey intentions more accurately, thus \name returns to question-answering interaction paradigm enriched with contextual information. }

\subsection{Contextual Computing for Interaction}

\Concise{Context-awareness can enhance interactive technologies' ability to discern user intentions. Advancements in wearable devices brought by integration of various sensors, exemplified by Pupil Labs Invisible glasses~\cite{tonsen2020high}, have enabled electronic devices to offer increasingly intelligent services.
}

\subsubsection{Gaze-based Intention Reasoning}

\Concise{Gaze has been applied across various tasks related to cognitive reasoning, including collaborative robotics~\cite{Li_2021-gaze-grasping, D'Angelo_2016-gaze-remote-collab, Andrist_2017bidirectional-gaze}, interactive technology inputs~\cite{Plopski_2023gaze-xr, Wei_2023target-selection, Pfeuffer_2015gaze-shifting}, reading assistant~\cite{gazereader, ajanki2009can}, cognitive load measurement~\cite{perkhofer2019using}, and intent classification~\cite{Newn_2019gaze-intentionAA, Doshi_2009driver-intention, Huang_2015intent-collaboration,xu2020recognizing}. Eye movements is a reliable indicator of user intent - ``eyes rarely visit objects that are irrelevant to the action''~\cite{land2006eye-movements-and-actions, tanriverdi2000interacting-with-eye}, which suggests a strong correlation between eye movements and inquiry intent in daily scenarios. Current methods for gaze-based intent prediction typically employ deep learning techniques, such as CNNs, for predefined classifications~\cite{admoni2016predicting-shared-autonomy, sattar2020deep-gaze-pooling}. Gaze data also serve as key indicator of interest in RBIR~\cite{ajanki2010contextual, Piening_2021looking-for-info}. To date, no research has integrated eye movements with natural language to deduce user intent.}

\subsubsection{In-situ voice-based interaction}

\Concise{Research in context-aware, command-based interaction systems has been extensively conducted. Studies on user programming of smart home appliances revealed a preference for ambiguous and multimodal command expression, such as combining gesture with spoken pronouns~\cite{liu2023understanding-in-situ, put-that-there}. To our best knowledge, there is a lack of research on both qualitative language patterns and quantitative pattern for gaze-enhanced voice interactions. This paper aims to fill this gap by presenting a user-enactment study leveraging \name's context-aware capabilities to identify patterns in user's in-situ expressions.}

\subsection{Multi-Modal Large Models}
Recent advancements in large language models (LLMs), including ChatGPT~\cite{chatgpt}, GPT-4~\cite{gpt4}, and others~\cite{llama,alpaca,vicuna2023}, have catalyzed research in multimodal information integration with substantial relevance to Human-Computer Interaction (HCI). These investigations can be primarily classified into two main approaches: Integrated Systems and Unified Multimodal Models. Each approach promotes fluid interaction between humans and computers, bridging vision and language to enhance user experience and efficiency in various applications.

\subsubsection{Integrated Systems Approach} 
This approach employs ChatGPT~\cite{chatgpt} as a central orchestrator, connecting specialized models designed for distinct visual tasks. Language prompts serve as bridges to invoke expert visual-language models, such as VisualChatGPT~\cite{visual_chat_gpt}, HuggingGPT~\cite{hugginggpt}, Cola~\cite{cola}, X-GPT \cite{zou2022generalized}, MM-REACT~\cite{yang2023mm}, and ViperGPT~\cite{suris2023vipergpt}. Despite offering a modular and adaptable solution for multimodal integration, this method presents challenges in individual model's training limitations and increased API query costs, potentially affecting the efficiency and scalability of HCI systems like \name{}.

\subsubsection{Unified Vision-Language Model Approaches}

\Concise{Efforts to integrate language models with visual processing have led to the creation of models like Flamingo, OpenFlamingo, and BLIP-2~\cite{flamingo, open_flamingo, li2023blip}. The release of GPT-4~\cite{gpt4} further catalyzed the development of advanced enterprise models including GPT-4 vision-language model (OpenAI), PaLM-E (Google), ERNIE (Baidu), Tongyi Qianwen (Alibaba), and SenseNova (Sensetime) ~\cite{gpt4, driess2023palm, ernie_bot, tongyi, sensenova}. In academia, contributions include LLaMA-Adapters, Mini-GPT4, LLaVA, Otter, and Voila-A~\cite{llama_adapter_v2, mini_gpt4, llava, li2023otter, yan2023voilaa}. LLaMA-Adapters extends LLaMA with an adapter module and multimodal prompts~\cite{llama}. Mini-GPT4 adapts BLIP-2's architecture, integrating Vicuna to replace the language decoder~\cite{li2023blip, vicuna2023}. LLaVA utilizes a trainable projector matrix for text-visual modality connections, which increases training demands~\cite{llava}. Otter, inspired by Flamingo, utilizes cross-gated attention layers for vision-language integration while freezing the vision encoder and language decoder to simplify training~\cite{flamingo, li2023otter}. Voila-A~\cite{yan2023voilaa} takes a further step,  discusses how to align vision-language models with user's gaze attention.}

\Concise{Despite these models are capable of engaging in image-related dialogues, they are limited by their dependence on users providing detailed queries and context. This reliance often leads to misunderstandings of user intent, thus limiting their practicality and real-world impact.}

\subsubsection{Multimodal Integration for Alignment} 

\Concise{Efforts to enhance fine-grained grounding and align with human intentions have incorporated diverse modalities, including bounding boxes~\cite{zhong2022regionclip,zhou2023regionblip,cho2021unifying,chen2023shikra}, patches~\cite{jin2023grill}, coordinate tokens~\cite{wang2022ofa,yao2022pevl,peng2023kosmos2}, and traces~\cite{yan2021control,pont2020connecting}. Despite advancements in multimodal integration, challenges persist in accurately interpreting context and intentions to provide pertinent, timely responses. This gap underscores the necessity for continued research to refine models that can adeptly navigate contextual nuances and human intentions, thereby enhancing the efficacy and intuitiveness of human-computer interactions as seen in \name{}.}

\section{The User-enactment study}

In this section, we first introduce the \name paradigm that combines gaze and voice inputs for information querying in daily scenarios. Then, we describe the prototype for an enactment experiment, which explores user's querying behavior under the \name paradigm.

\subsection{Introducing \name: A Gaze-Facilitated Querying Paradigm} \label{sec:intro_voila}
\textbf{We propose \name as a novel ubiquitous querying paradigm that combines users' gaze data, visual field and voice-based natural language question as input.} The design concerns of \name can be summarized in three folds:

\begin{enumerate}
    \item \textbf{Hands-free:} \name is designed to be deployed on wearable devices with gaze sensors, such as smart glasses and VR/AR systems, allowing users to post questions verbally without the need to operate the device with hands.
    \item \textbf{In-context:} \name's ability to capture visual field provides environmental clues relevant to users' queries, and in turn, users may formulate questions closely related to their current situation. 
    \item \textbf{Multi-modality:} \name enables multi-modality data inputs that can complement each other to produce a more comprehensive understanding of users' questions. Consequently, users can express queries in a more flexible manner.
\end{enumerate}

\Concise{In traditional mobile device usage, users typically focus on relevant context in their surroundings before inputting necessary data into the application. However, given \name's properties, users can express in multiple modalities simultaneously, unlike mobile systems that necessitate sequential input of natural language or media.} This raises a unique research question: do users' gaze patterns show consistency with their spoken natural language, and if so, how do they relate spatiotemporally?

\subsection{The User-Enactment Study}
To investigate user's querying behavior with \name, we conducted a user-enactment study where participants submit queries in any manner they believed \name should be able to comprehend and respond to, given its sensing capabilities.

\subsubsection{Scenarios and Participants.} 
Each participant partake in one of three daily scenarios: supermarket shopping, museum visits, and domestic living. For supermarket and museum, we selected publicly accessible and popular locations within local community. For domestic living, we established an independent room furnished with at least 10 pieces of furniture and 20 common household items. We enrolled 21 participants (8, 5, 8 for each scene), consisting of 13 males and 8 females aged from 19 to 30 ($std=3.18$). The duration of participation ranged from 90 to 150 minutes, with a compensation rate of 30 USD per hour.

\subsubsection{Hardware Setup.}

\Concise{The Pupil Labs Invisible~\cite{tonsen2020high}~\footnote{The Invisible glasses rely on a connection to a mobile phone to be driven; thus we advised participants
to place the phone in their pockets to minimize potential distractions.} is a pair of gaze-tracking smart glasses widely used in research. Featuring gaze sensors, egocentric camera, and microphone, it meets the needs for \name. Participants were instructed to wear the glasses without any headdress that might cause occlusion to the sensors. Data was continuously recorded as participants engaged in their chosen scenario.}

\subsubsection{Procedure.} \label{sec:exp1_procedure}
The study comprises three sessions: an introduction session, an enactment session where participants pose queries within their chosen scenario, and a discussion session for reviewing their queries and exploring insightful queries through dialogue with the experimenter.

\Concise{The introduction session aims to ensure that participants understood \name's sensing capabilities and encourage them to come up with diverse query formats. We provided an analogical script of conversation with friends, illustrating how multiple expression modalities can be employed in daily life. Then participants put on the Invisible glasses and viewed a real-time egocentric video with overlaid gaze trace, which facilitated their understanding of the gaze-tracking.  Each scenario was given a distinct purpose: purchasing ingredients for a lavish dinner, exploring an exhibition for intriguing knowledge, and spending a day at home or working remotely. This general purpose was assigned to prevent participants from feeling aimless and confused in the scene, since they wouldn't receive further instructions or interaction with the experimenter during the enactment session. We also briefed participants on subsequent study procedures.}

\Concise{In the enactment session, participants were required to wear the Invisible glasses and engaged with their chosen scenario. They were encouraged to ask any questions related to the scene. Participants did not receive any responses during this session. A minimum of 10 questions was required. All sensors' data were recorded throughout this session. }

\begin{figure}
    \centering
    \includegraphics[width=0.9\linewidth]{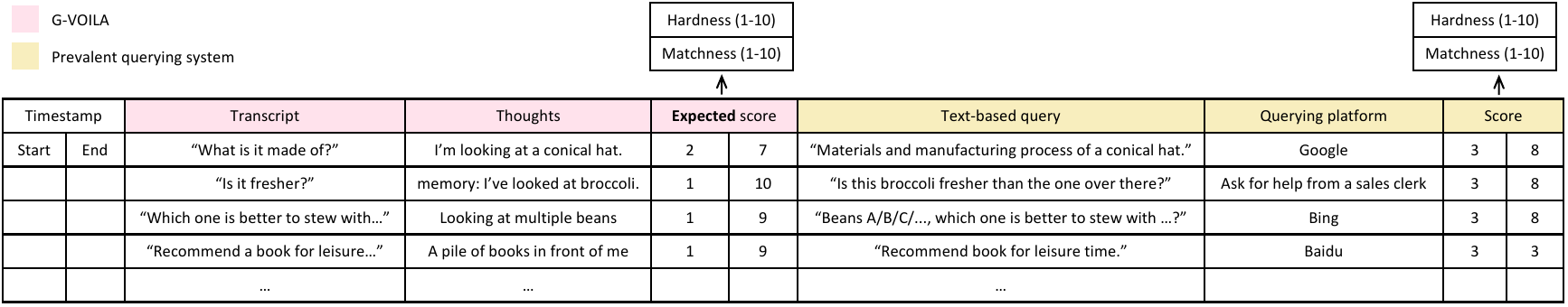}
    \caption{Discussion chart for the user-enactment study. Headers and four illustrative examples were shown in the chart. For the first example, participant supposed \name could understand ``it'' to be a conical hat because (s)he was looking at it, whereas there is no clue showing that (s)he also wanted to know the manufacture process, thus \name should result in a lower matchness score. \RIII{But when querying with her daily used platform google, she might rephrase her question to be more clarified.}}
    \label{fig:exp1_chart}
\end{figure}

\Concise{In the discussion session, participants reviewed the recorded video (overlaid with gaze traces) from the enactment session and completed a form for each query, as depicted in Figure~\ref{fig:exp1_chart}. For each query, they reformulated their queries to search with prevalent IR systems until they received satisfied information or encountered frustration.} \RIII{Based on their daily problem-solving preferences, participants can choose from digital search tools to human interactions (consulting sales clerks) as their querying platform.}

\Concise{For each query, Participants rated (1) the hardness in finding desired answer, including query formulation and information scanning, and (2) the matchness of the information retrieved. The final query text and utilized querying systems were documented. Although \name could not respond during the study, participants still evaluated it in terms of hardness and matchness. Matchness scores were assigned based on participants' perceptions of whether \name could comprehend their verbal question and provide a satisfactory response, given its sensing capabilities.}

\Concise{After reviewing all queries, the experimenter identified queries where significant discrepancies exist between the verbal and textual queries, discussing with participants about how they assumed \name could understand and respond to such question. Notable insights were recorded in the ``thoughts'' column.} Timestamps were later added to the form by the experimenter.

\section{Analysis and Findings} \label{sec:findings}

In three predetermined daily life scenarios, we collected a total of 418 valid inquiry data and annotation pairs. The collected data was analyzed to comprehend users' multi-modal inquiry form-factors within the \name paradigm. As discussed in Section~\ref{sec:intro_voila}, we aimed to answer three research questions:

\begin{enumerate}
\item \textbf{How do users perceive and anticipate an IR system under the \name paradigm?}
\item \textbf{What distinctive language expression patterns do users exhibit with \name?}
\item \textbf{How are users' gaze patterns spatiotemporally related to their spoken query language?}
\end{enumerate}

\subsection{User's Comprehension and Expectation of \name} \label{sec:expectation}

\begin{figure}
  \centering
  \subcaptionbox{Category counts\label{fig:exp1_catestat}}
    {\includegraphics[width=0.45\linewidth]{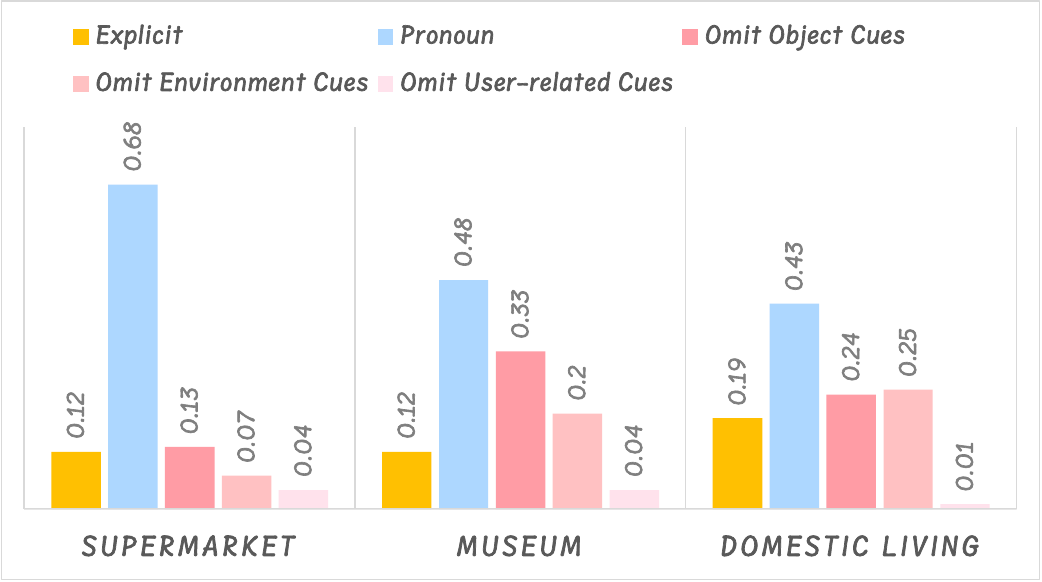}}
  \subcaptionbox{Hardness and Matchness\label{fig:exp1_scorestat}}
    {\includegraphics[width=0.2\linewidth]{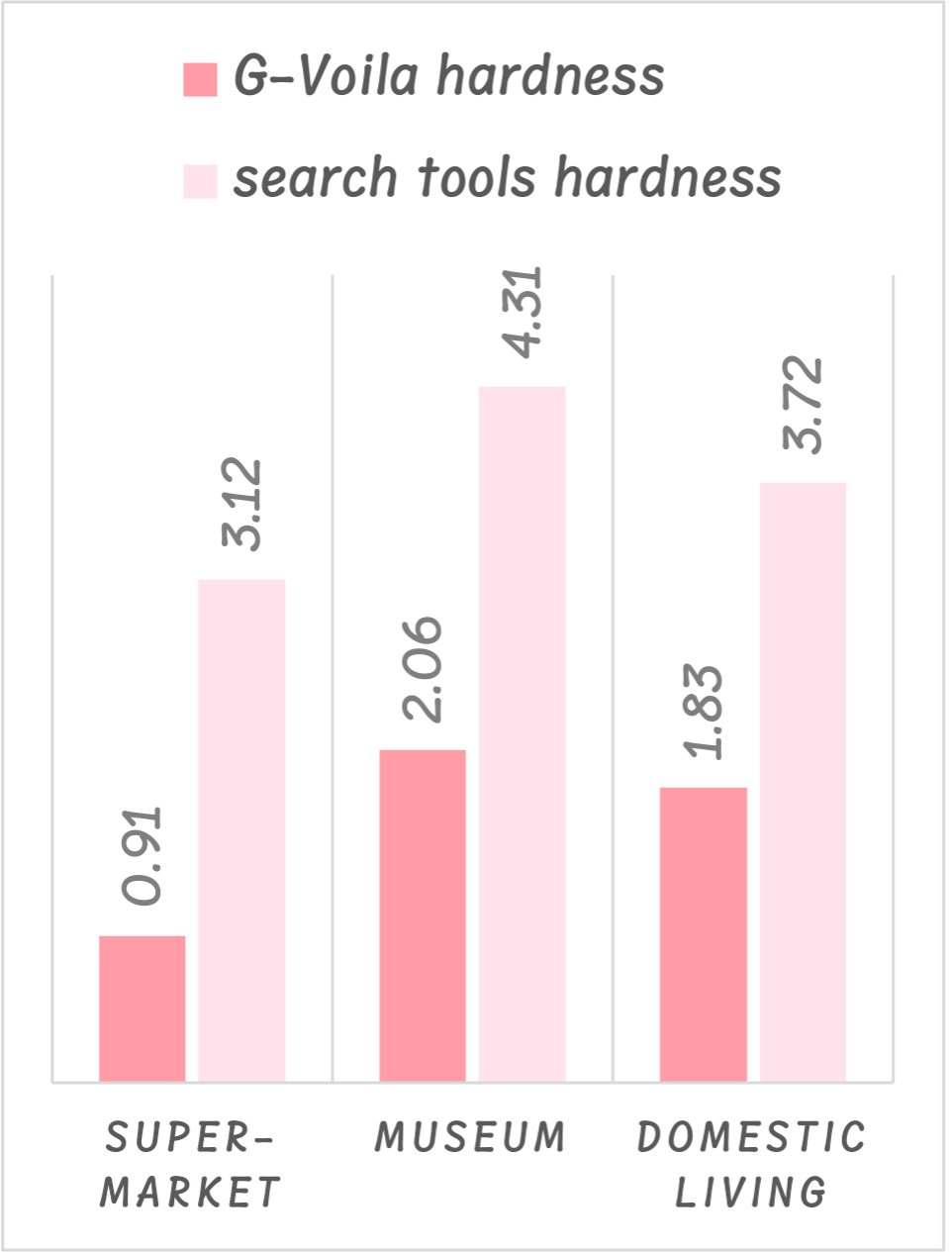}
    \includegraphics[width=0.2\linewidth]{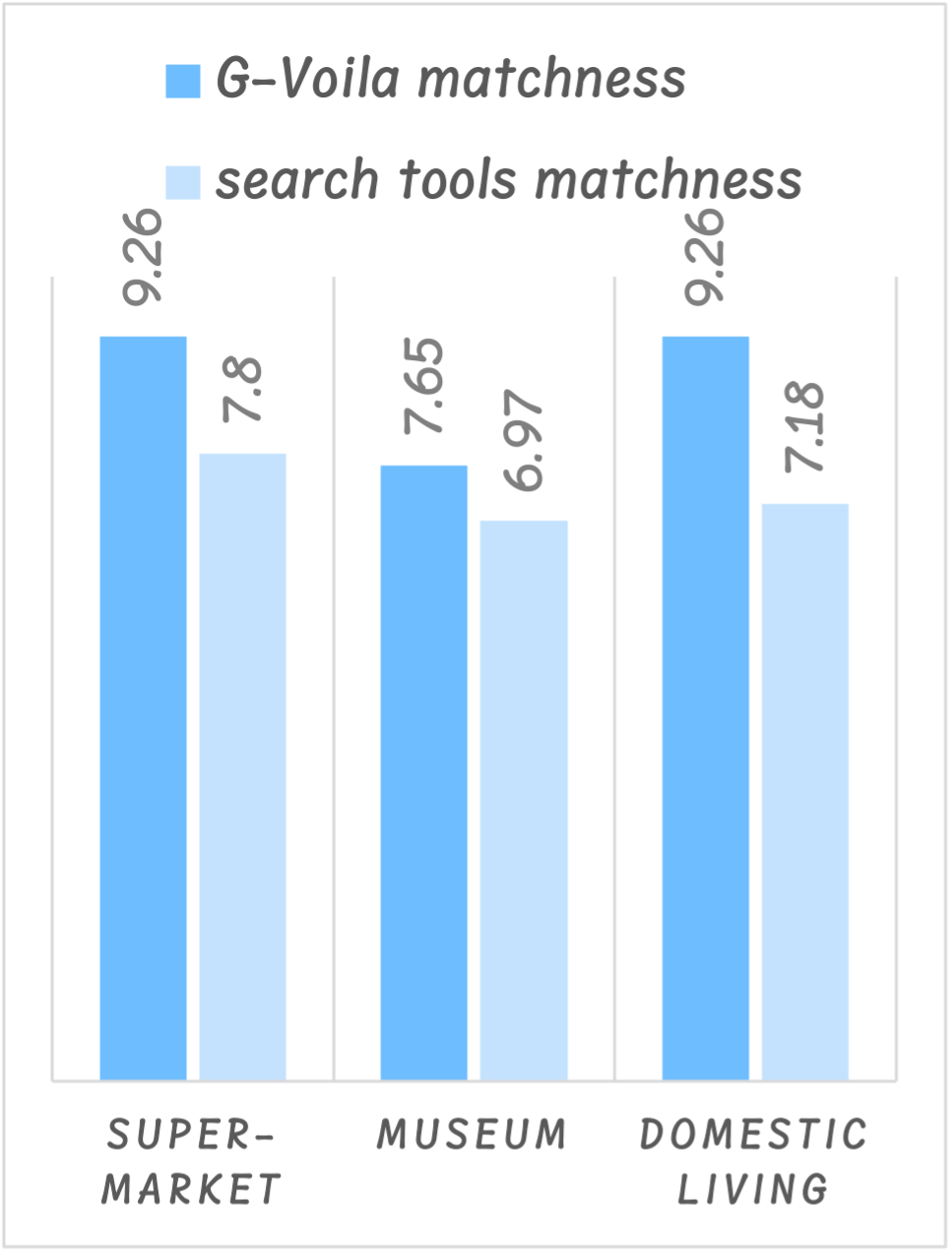}}
  \caption{Statistic results for the formative study, shown for all scenarios. (a) A bar plot displaying statistic count of queries for each category. \RIII{All queries are classified it's ambiguity, which taxonomy is presented in Section~\ref{sec:language_pattern} and Figure~\ref{fig:exp1_category}}. \RIII{(b) Under the user-envisaged \name{} and text-based IR approaches, the hardness of seeking solutions and answer's matchness to user's query intent.}}
  \label{fig:exp1_stat}
\end{figure}
\Concise{To understand users' comprehension and expectations of \name, we computed average hardness and matchness scores for each scenario, as shown in Figure~\ref{fig:exp1_scorestat}. Lower hardness and higher matchness scores suggest users anticipate \name to better interpret casual queries. Qualitative analysis of the ``thoughts'' content yielded three key aspects for \name to enhance its functionality:
}

\begin{enumerate}
    \item \textbf{User interests indicated by gaze content:} When users do not provide sufficient descriptions or indications of their query objects, \name should be capable of extracting this information by examining users' gaze data.
    \Concise{\item \textbf{Contextual information derived from visual field:} \name should infer situational context and answer constraints from the shared visual environment when user descriptions are insufficient.}
    \Concise{\item \textbf{Incorporating features from other systems:} Users expect \name to incorporate features characteristic of other systems they have experienced, such as controlling appliance or memorizing personal preferences. }
\end{enumerate}

\begin{figure}
    \centering
    \includegraphics[width=0.9\linewidth]{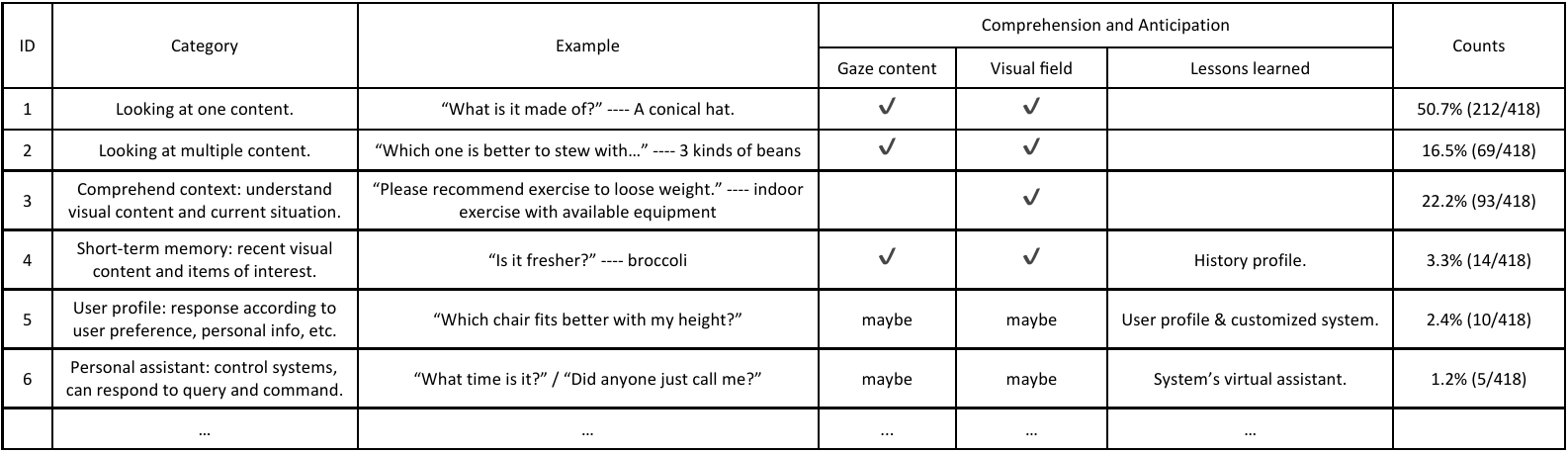}
    \caption{Breakdown of user queries by their anticipation of \name usage. ``Lessons learned'' are short for ``Lessons learned from other well-designed system''. At least one query example are provided for each category.}
    \label{fig:exp1_anticipation}
\end{figure}

As depicted in Figure~\ref{fig:exp1_anticipation}, user queries were classified according to their anticipation of \name's functionality. The majority of queries (67.2\%, n=281) requires \name to enrich responses by interpreting users' eye movements to identify objects of interest, focusing on either single (50.7\%, n=212) or multiple objects (16.5\%, n=69). The shared visual field also serves as another crucial input stream (22.2\%, n=93) by providing either the context of the activity (e.g., "can I drink it now" in a supermarket) or the surroundings layout (e.g., "where to place my glasses" at home). Fewer queries (6.9\%, n=29) draw on insights gained from other interaction systems, involving elements like history and user profiles or operating system assistance. These occasionally depend on \name's sensing ability, such as constructing a reliable and privacy-aware history profile based on gaze and egocentric views.

\takeaways{
\RIV{\textit{Take-aways:} The majority of user queries anticipate \name to leverage gaze data to pinpoint objects of interest (67.2\%) and visual data to supplement context information (22.2\%).}
}

\subsection{User's Verbal Expression Patterns with \name} \label{sec:language_pattern}

\begin{figure}
    \centering
    \includegraphics[width=0.9\linewidth]{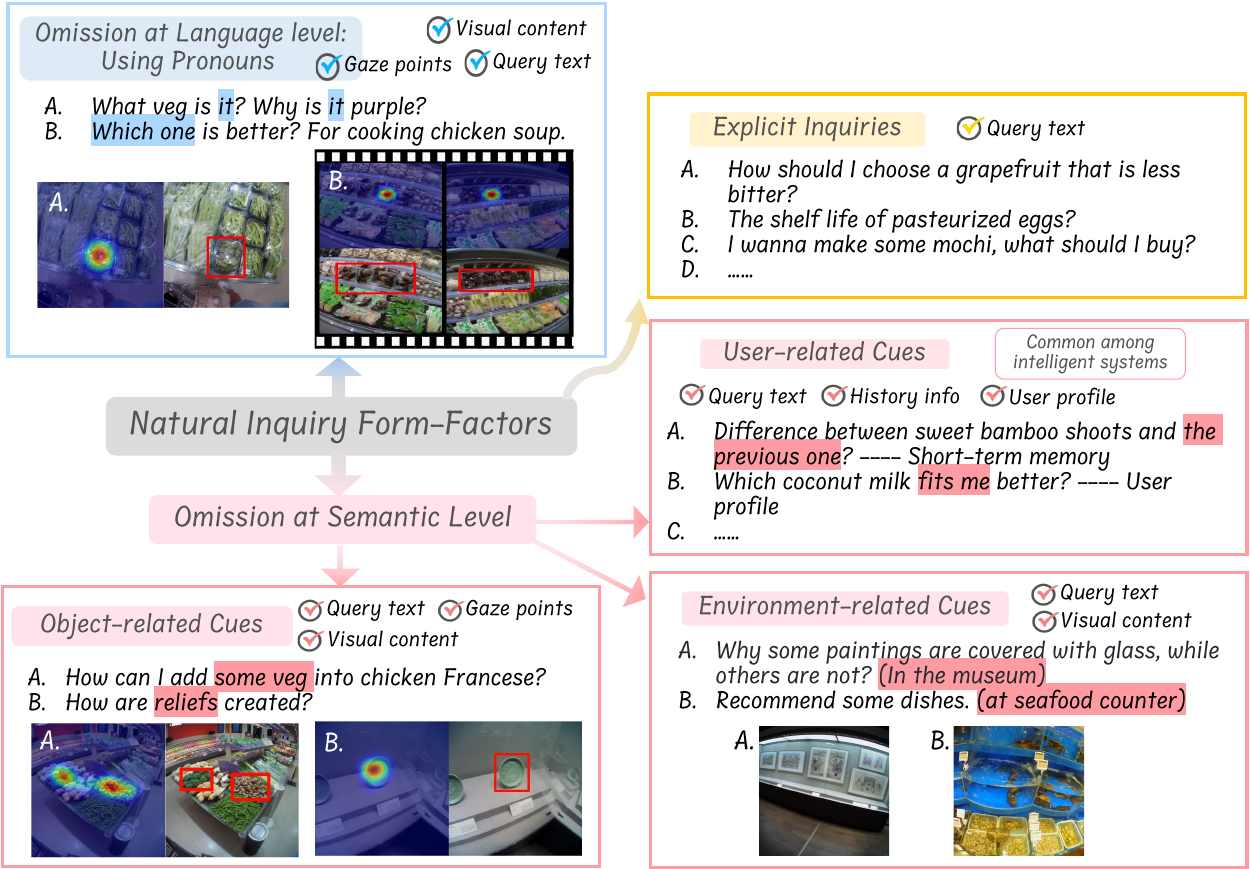}
    \caption{Exploring Inquiry Form-Factors with \name{}. The identified categories were shown in each card, with a checkmark indicating the required data for refilling user's query intent. Keywords in the query text are highlighted, and representative frames are selected from each query video.}
    \label{fig:exp1_category}
\end{figure}

\Concise{We compared user's verbal questions with their reformulated queries textto discern language expression patterns with \name, as illustrated in Figure~\ref{fig:exp1_category}. We identified the following categories considering the diction and semantic gaps:}

\Concise{\begin{itemize}
    \item \textbf{Explicit Inquiries.} Intelligent assistants could understand and respond to inquiries using only the verbal question and common knowledge. Natural language's ambiguity and the discrepancy between spoken and written expression present challenges, which can be mitigated by LLMs as evidenced by prior research.
    \item \textbf{Omission with Language Indicator -- Using Pronouns.} Intelligent agents could identify the existence of ambiguity by examining the query text, often indicated by the usage of pronouns. By replacing the pronouns with referred objects, the agent can convert it into an explicit inquiry. Gaze data can clarify the referred objects, although deciphering a user's gaze pattern to pinpoint when they begin to look at the targeted items remains challenging.
    \item \textbf{Omission at Semantic Level.} Queries may seem complete linguistically but lack specific descriptive information, leading to a broader answer scope. Including:
    \begin{itemize} 
        \item Object-related cues. Such omission may be discovered by aligning gaze content and query text, e.g. ``relief on porcelain plates'' and ``some veg'' referring to specific kinds of vegetables.
        \item Environment-related cues, where lack of context may yield answers unsuited to the current setting, e.g., suggesting painting placement in a museum instead of an art classroom or other locations.
        \item User-related cues. Since \name's uniqueness lies in its gaze sensing and shared visual field, we designated other types of omissions that cannot be compensated by gaze and visual content as ``user-related cues'', such as historical information or user preferences.
    \end{itemize}
\end{itemize}}

In accordance with this taxonomy, three researchers examined each query record and reviewed the corresponding video clips (overlaid with gaze data) to classify all queries, eventually reaching a consensus. Figure~\ref{fig:exp1_catestat} displays the statistical results for each category.

\takeaways{
\RIV{\textit{Take-aways:} The majority of user queries involve usage of pronouns (68\%, 48\%, 43\%) and omission of certain constraint (20\%, 40\%, 39\%).}
}

\subsection{Gaze Patterns and their Spatiotemporal Relationship to Spoken Query Language} \label{sec:gaze_pattern}

\begin{figure}
  \centering
    \subcaptionbox{around staring\label{fig:max_fixation}}
    {\includegraphics[width=0.4\linewidth]{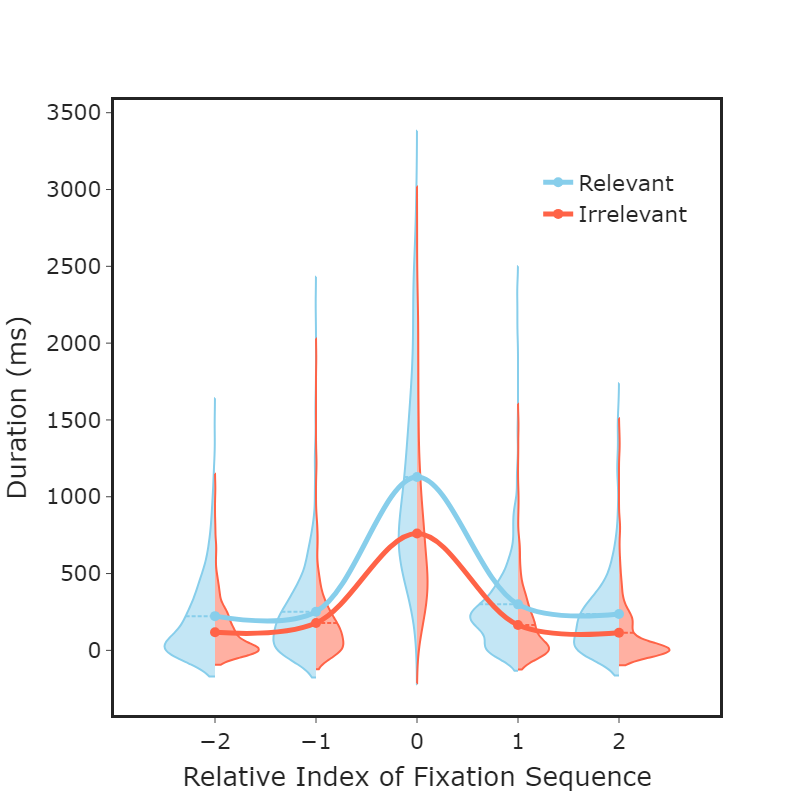}}
    \subcaptionbox{around pronouns\label{fig:fixation_pronoun}}
    {\includegraphics[width=0.42\linewidth]{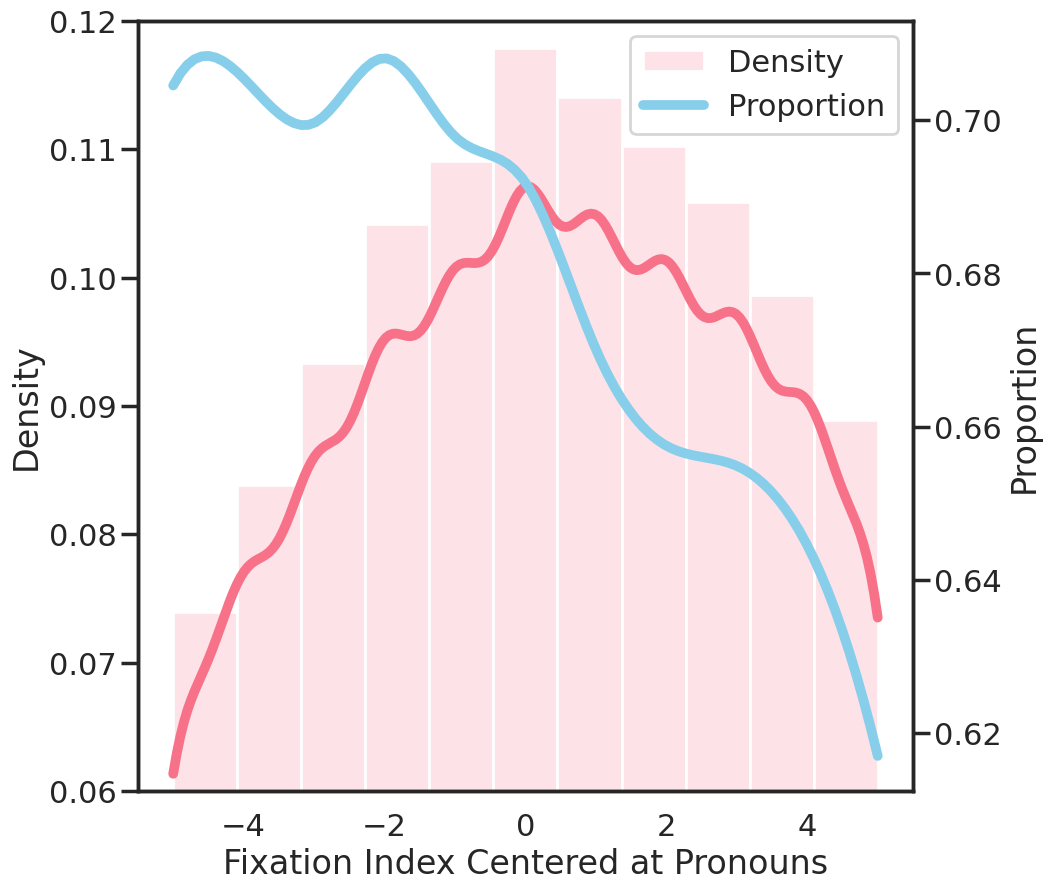}}
    
  \subcaptionbox{absolute distribution\label{fig:relevant _distribution}}
    {\includegraphics[width=0.45\linewidth]{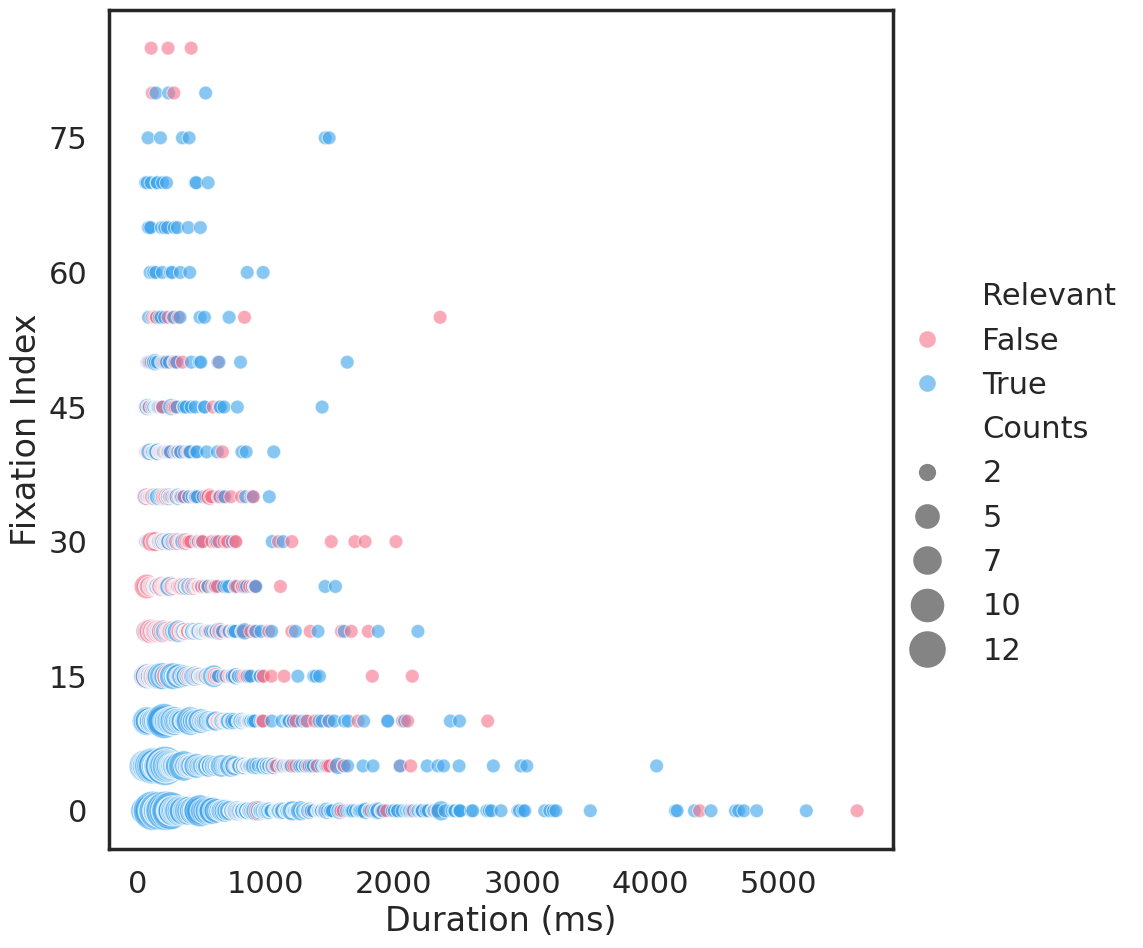}}
    \subcaptionbox{distribution of startup time\label{fig:before_querying}}
    {\includegraphics[width=0.39\linewidth]{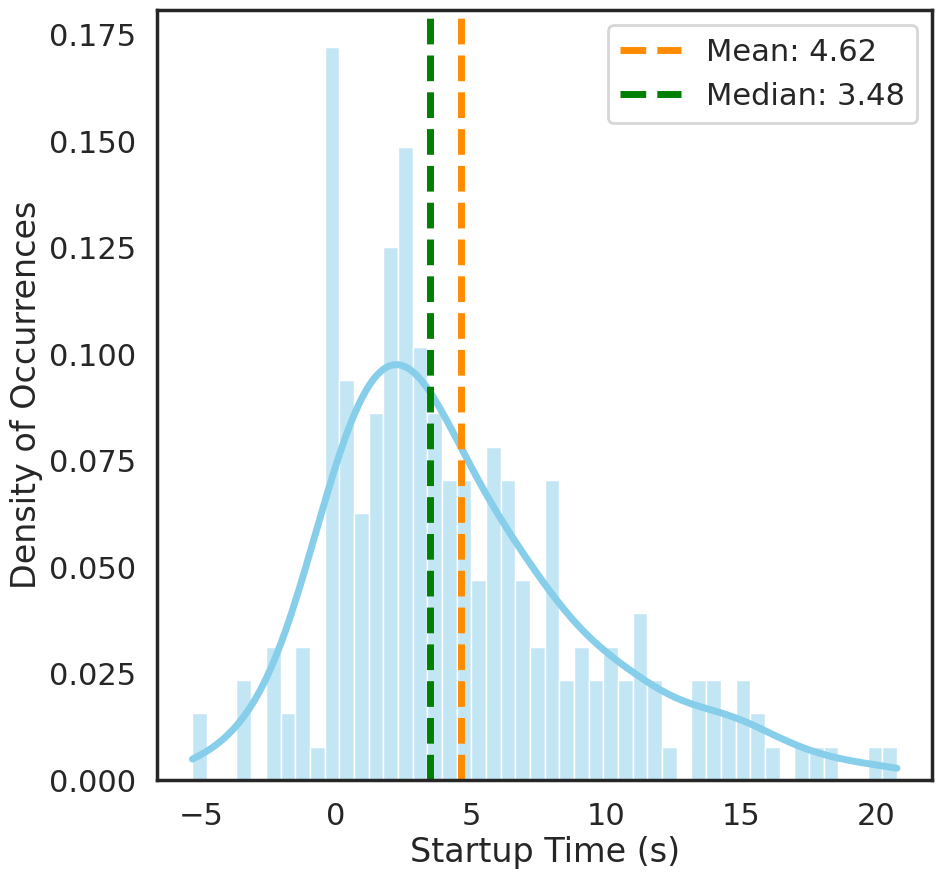}}
  \caption{Relevant fixation distribution. (a) A bar plot and approximation curve illustrating relevant fixation density centering around pronoun during querying. (b) A violin plot comparing duration of relevant and irrelevant fixation peak. (c) A scatter plot demonstrating fixation's occurrence order during querying. \RIII{(d) A distribution plot of startup time for all querying.}}
  \label{fig:exp1_analysis}
\end{figure}

We conducted quantitative analysis on the spatiotemporal correlation between users' gaze patterns and their spoken questions, revealing a mouth-eye coordination when querying with \name, as depicted in Figure~\ref{fig:exp1_analysis}.

\subsubsection{Data preprocessing.}
We leveraged fixation annotations $F: \{f_n\}$ provided by Pupil Labs, where each fixation $f_i$ consists a tuple of $(t_{start}^i,t_{end}^i,x_i,y_i)$, representing the start and end times, and the fixation coordinates. Each spoken query $q\in Q$ is represented by $(t_{start}^q, t_{end}^q, W_q)$, with $W_q = \{w_{q1}, w_{q2}, ..., w_{qn}\}$ denoting the sequence of spoken words. Time segments $T_q$ for each word $w_{qj} \in W_q$ were generated using~\cite{whisper-timestamped, radford2023robust, JSSv031i07}, with $t_{qj}: (t_{start}^{qj}, t_{end}^{qj})$ specifying the timing of each word. 

Fixations were labeled with it's gaze content by manually inspecting video clips (overlaid with gaze trace and fixation points). We established a binary mapping function $g:F\times Q\rightarrow\{ 0, 1\}$, where $g(f_i, q)=1$ if the gaze content of $f_i$ is relevant to query $q$, and $g(f_i, q)=0$ otherwise. Relevancy was determined by whether the fixation's content matched the expected answer or was mentioned in the query.

We mapped each word to sequential fixations using $a: W\rightarrow F$, where $f_i \in a(w_{qj})$ if $[t_{start}^i,t_{end}^i] \cap [t_{start}^{qj}, t_{end}^{qj}] \neq \varnothing$. This function indicates the concurrent of word $w_{qj}$ and fixation $f_i$.

\subsubsection{\textbf{Users tend to focus more on relevant objects than irrelevant ones when conveying queries.}} \label{sec:gaze_relevancy}
We compared the longest relevant fixation to the irrelevant one. For each query $q$ with fixation sequence $\{f_n\}$, we identified the indices of the longest relevant fixation $f_i$ and the longest irrelevant fixation $f_j$, i.e. $$
i = \argmax_i \{t_{end}^i - t_{start}^i \mid g(f_i, q) = 1\}, j = \argmax_j \{t_{end}^j - t_{start}^j \mid g(f_j, q) = 0\}$$
Both $i$ and $j$ were denoted as relative zero index, then the distribution and mean duration of adjacent fixations were plotted in Figure~\ref{fig:max_fixation}. 

The results show that relevant fixation are significantly longer than irrelevant ones,  indicating user's propensity to concentrate on targeted objects rather than unrelated surroundings. This finding supports the idea that the increased focus on relevant objects may facilitate better understanding and communication of the query, as well as potentially indicate the user's cognitive processing in relation to the question being asked.

\takeaways{
\RIV{\textit{Take-aways:} Although users may have wandering gaze on irrelevant surroundings when expressing queries, they tend to focus longer on relevant objects($mean\approx 1.1s$) than irrelevant objects ($mean\approx 0.75s$).}
}

\subsubsection{\textbf{Co-occurrence exists between using pronouns and looking at referred items.}} \label{sec:gaze_around_pronoun}

Pronouns $w_{qj}$ were located within each query's word sequence $W_q$, with $j_{pron}$ representing the pronoun's index. For each query's pronoun, we quantified relevant fixations on nearby words (within a window of $r \in [-5, 5]$ words) as $$c(r) = Count\{f_i \mid f_i \in a(w_{q(j_{pron}+r)}), g(f_i,q)=1, q \in Q\}$$

where $r$ is the relative word position. Based on the calculation of $c(r)$, the density of relevant fixations with relative indices ($r$) was plotted as the ``density'' distribution in Figure~\ref{fig:fixation_pronoun}. Additionally, we calculated the proportion of relevant to total fixations around each pronoun as follows, and presented as the ``proportion'' plot in Figure~\ref{fig:fixation_pronoun}. $$p(r) = {c(r) \over c_{all}(r)}, c_{all}(r) = Count\{f_i \mid f_i \in a(w_{q(j_{pron}+r)}), q \in Q\}$$

The ``Density'' plot peaks at the pronoun's index (relative zero index), indicating a higher occurrence of relevant fixations when users are speaking the pronoun. Conversely, the ``Proportion'' plot shows a decline following the spoken of pronoun. This suggests a shift in focus away from the referent to the next objects of interest or irrelevant stimuli, which happens right after the pronouns are spoken. This pattern may reflect a transition in cognitive focus post-referent identification.

\takeaways{
\RIV{\textit{Take-aways:} User's visual attention on relevant objects peaks at the mention of pronouns and then shifts away, indicating the existence of mouth-eye coordination but with gaze movements ahead.}
}

\subsubsection{\textbf{Users tend to look at objects that are relevant to their query content at the beginning of the query.}}
We draw a distribution plot of fixation relevancy $g(f_i, q)$ for all queries, in terms of its fixation index $i$ and duration $t_{end}^q - t_{start}^q$, as illustrated in Figure~\ref{fig:relevant _distribution}. As observed from the plot, blue dots representing relevant fixations dominate the lower left corner and tend to have larger radius at lower sequential index. This indicates a concentration of relevant fixations with longer durations in the initial stages of query formulation, revealing the fact that users are more focused on relevant objects when starting a query. The term ``focus'' here can be interpreted as either longer duration or higher frequency.

This trend aligns with the concept that ``gaze often moves on before the last act is complete''~\cite{land2006eye-movements-and-actions}, implying that user's attention is more focused at the onset of a task. Furthermore, considering the next act, gaze moving on before the last act is complete means that gaze may move to the next act before it starts, which inspired us to ask the research question: Do users tend to look at items of interest before starting to ask questions? If so, how long? 

\takeaways{
\RIV{\textit{Take-aways:} Users demonstrate a pronounced focus on relevant objects at the beginning of query formulation.}
}

\subsubsection{\textbf{Users tend to look at items of interest before starting to ask questions.}}
We introduce the concept of ``startup time'' to denote the interval during which users visually engage with their query subject before verbalizing the question. Upon analyzing adjacent query intervals, we established a maximum threshold of 20 seconds for the startup time. For each query $q$, the startup fixation $f_{i_s}$ and corresponding startup time $t_{startup}^q$ are determined by: $$i_s = \argmax_i \{ t_{start}^q - t_{start}^i \mid t_{start}^q - t_{start}^i \le 20, g(f_i, q)=1\}, t_{startup}^q = t_{start}^q - t_{start}^{i_s}$$

Figure~\ref{fig:before_querying} displays the distribution of startup times across all queries, highlighting an average startup time of 4.62 seconds and a median of 3.48 seconds. We further identified a behavior of ``turning back'', where users briefly disengage from an object and then redirect their attention back to it when formulating a related query.

\takeaways{
\RIV{\textit{Take-aways:} Users typically engage with objects of interest before initiating a related query, with an average start up time of 4.62s.}
}

\section{Design Framework for \name Paradigm}

\begin{figure}
    \centering
    \includegraphics[width=0.5\linewidth]{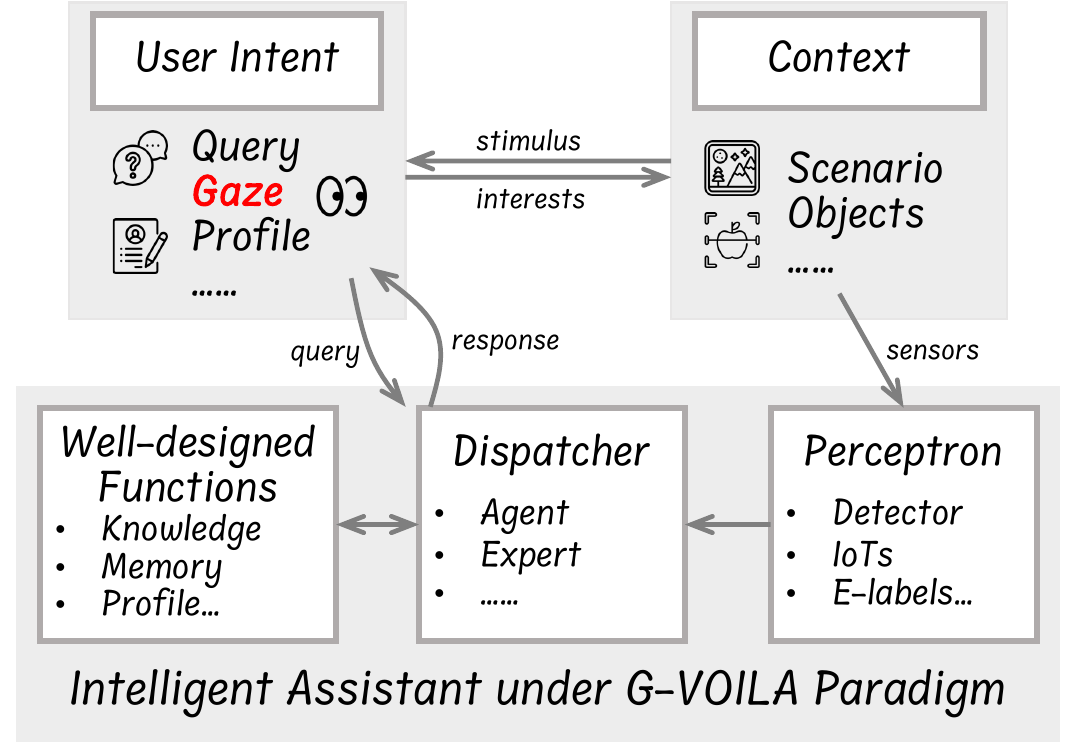}
    \caption{A concept map illustrating the information flow between user, environment, and \name{} during one query.}
    \label{fig:model_thought}
\end{figure}

In this section, we proposed a design framework for intelligent assistants within the \name{} paradigm. Our design framework tackles critical issues of incorporating gaze data into the query response process, considering (1) spatial indicators, (2) temporal relationships to the query expression, and (3) semantic connections between gaze content and query intent.

Building upon the user comprehension and expectations of \name (Section~\ref{sec:expectation}), as shown in Figure~\ref{fig:model_thought}, we have delineated a modular envision for intelligent systems within the \name paradigm. This encompasses three core elements: the user, the surrounding context, and the system itself. The interaction procedure of conducting a single query can be decomposed as follow: triggered by a specific scenario, the user manifests interest in particular contextual information and subsequently initiates a query to the intelligent assistant; given the propensity for ambiguity in user's language expressions (Section~\ref{sec:language_pattern}), the intelligent assistant is tasked with deducing user's explicit query intention by leveraging information perceived from gaze trackers and egocentric camera (Figure~\ref{fig:exp1_anticipation}, Section~\ref{sec:gaze_pattern}); meanwhile, the user’s query may necessitate the activation of specialized functionalities (Section~\ref{sec:expectation}), where the intelligent assistant should dispatch the appropriate functions to efficiently address the query; by then, a satisfactory response can be delivered to the user.

\begin{figure}
    \centering
    \includegraphics[width=\linewidth]{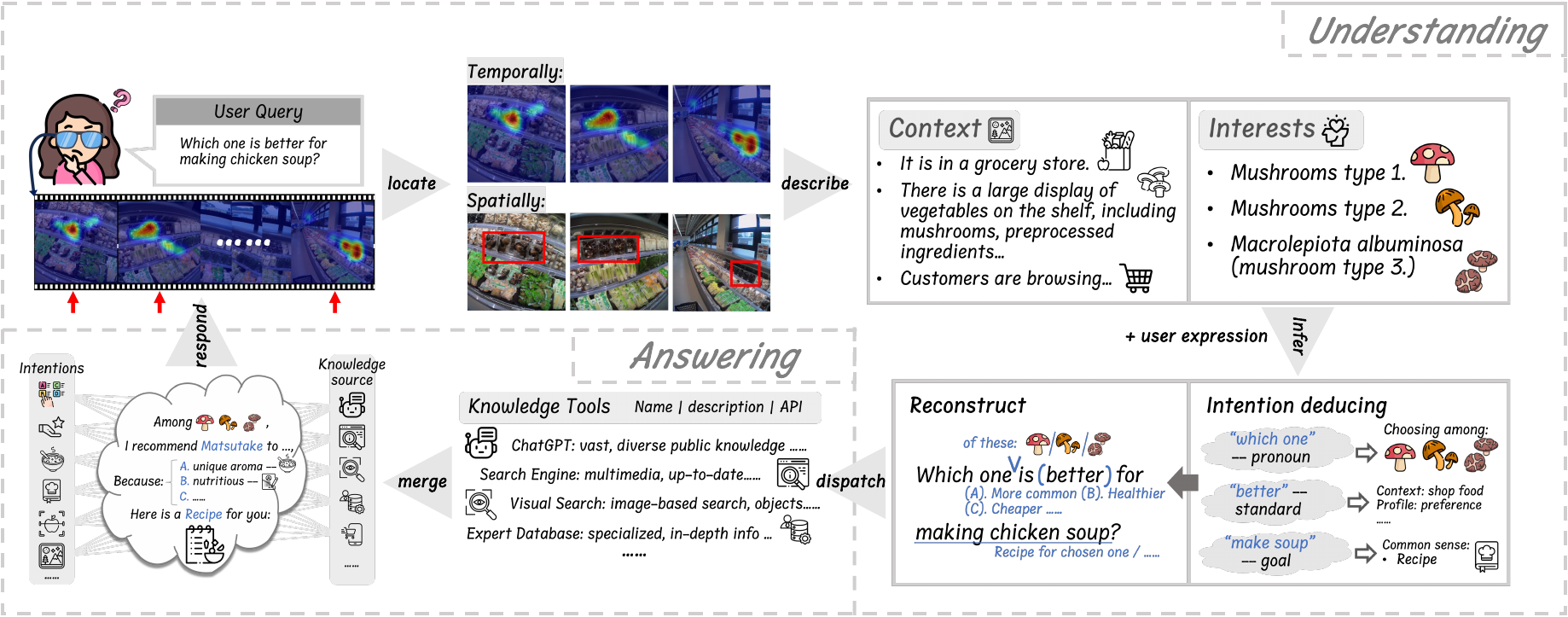}
    \caption{\name{}'s inference and generate pipeline. Each step is indicated by a gray arrow and label.}
    \label{fig:pipeline}
\end{figure}

The information divergence between the user and the intelligent assistant primarily stems from the assistant’s lack of insight into (1)\textbf{when} the user become interested, (2)\textbf{what} sparked their interest, and (3)\textbf{how} their curiosity is specifically oriented. Hence, addressing this gap necessitates a joint analysis of \name's multi-modal sensory inputs to accurately deduce the user's query intent, which should be consistently reflected throughout the response mechanism. In light of this, as shown in Figure~\ref{fig:pipeline}, we advocate a novel design principle for IR systems within the \name paradigm, which partitions the response process into two distinct phase: initially, multi-modal inputs is leveraged to discern and augment the user's intent; subsequently, responses are generated meticulously aligned with the refined understanding of the user's intent. Details will be discussed in the upcoming subsections.

\subsection{Gaze-Facilitated: Deciphering In-Situ Query Intention}

Contemporary text-based Information Retrieval (IR) techniques, including search engines and Large Language Models (LLMs), have demonstrated robust proficiency in addressing explicit query texts and mitigating ambiguities inherent in natural language. However, within the \name paradigm, \textbf{a novel challenge emerges: how to explicate user's ambiguous query expression through the integration of gaze and visual data?} To clarify the framework for deciphering in-situ query intentions, we dissect this process into three distinct components, each elucidated in detail. While these components are presented as separate entities for conceptual clarity, they can be, in practical applications, implemented by a single multi-modal model.

\subsubsection{\textbf{Temporal and Spatial Localization of Interest}} \label{sec:pipeline_step1} 
The data stream within the \name paradigm can be conceptualized as a chronological series of video and eye gaze data. This necessitates a meticulous cross-referencing approach, both temporally and spatially, to pinpoint elements within the three-dimensional scope of the video data (time dimension and two-dimensional image frames) that are pertinent to the user’s query intent.

Discussion in Section~\ref{sec:gaze_pattern} underscores the significance of temporal localization: \textbf{users are not always looking at their intended focus; instances of wandering gaze or moving on in advance are commonplace.} A poorly devoted temporal localization strategy risks emphasizing video frames that correspond to irrelevant gaze points. Such a misalignment can lead to a disastrous deviation from the user's original intent, given that gaze coordinates should be leveraged for spatial localization. Conversely, one distinctive feature of \name usage is the synchronicity between eye movements and verbal expressions. As corroborated in Section~\ref{sec:gaze_around_pronoun}, \textbf{temporal alignment exists between the semantics of gaze content and the semantics of the user's spoken words}, which indicates a robust correlation between gaze and voice modalities. To conclude, the selection of the appropriate data frames for intent analysis should consider gaze pattern's intrinsic nature, as well as can be potentially guided by the timeline of spoken words.

Although gaze coordinates can indicate the spatial location of a user's interest (Section~\ref{sec:gaze_relevancy}), they fall short in predicting the interest range. This limitation becomes particularly salient in scenarios where multiple objects are spatially nested. In the case where user's gaze is directed at a flower within a vase on a shelf, the specific object of interest becomes equivocal: the flower solely, the flower-vase ensemble, or the shelf. Our user-enactment study occurred several instances of such nested structures within the user's field of vision. A notable example involved a user standing in front of a pile of water tank filled with all sorts of seafood and querying about a specific kind of fish.

To address this ambiguity, we propose two indicative cues for disambiguating the precise range of interest. \textbf{The first cue is the user's linguistic expressions, which can used to align with the semantic of items at different granularity.} For example, a request like ``recommend a dish for this'' would most likely pertain to a specific type of fish, whereas ``recommend a dish'' might refer to either a category of seafood or a specific variety. \textbf{The second cue is enhanced robustness by analyzing the gaze trace over a brief duration.} In this context, a broad sweep across the water tank would suggest an interest in the general category of seafood, while a relatively steady gaze on a specific section might indicate interest in a particular fish variety.

\subsubsection{Semantically Understand Visual Context and Gaze Content}
In the pursuit of offering a precise response to a user's query, a system must primarily grasp the visual context, encompassing elements like the background and location. One effective approach to achieving this is by imbuing descriptions with an appropriate level of semantic understanding. Simultaneously, this necessitates the system's capability to differentiate between coarse and fine levels of semantic granularity. A coarse semantic description might simply categorize an object as a ``plate'', while a fine-grained description would delve into its material, color, pattern, and distinctive attributes, such as labeling it as a ``ceramic plate adorned with a relief of a dragon''. The depth of semantic understanding should harmonize with the nature of the question being addressed. In essence, this constitutes a multimodal understanding challenge that is independent of our paradigm design. Every enhancement in this domain can seamlessly integrate into our paradigm, resulting in improved performance.

\subsubsection{\textbf{Deduce Intention and Reconstruct Query Expression}} \label{sec:pipeline_step3}
Guided by the taxonomy established in Section~\ref{sec:language_pattern}, query expressions that inadequately convey the user's intended meaning typically suffer from two deficiencies: the utilization of ambiguous terms, such as pronouns, or the exclusion of pivotal information with no discernible hints in the expression. Addressing the former necessitates the substitution of ambiguous language with contextual information acquired from alternative modalities, as elaborated upon in Sections~\ref{sec:gaze_relevancy} and \ref{sec:gaze_around_pronoun}, with a particular focus on gaze data. Conversely, resolving the latter challenge involves a strategic integration of additional information, sourced either from perceptual inputs or user-specific contexts. For both cases, the deployment of a meticulously formulated mechanism described in Section\ref{sec:pipeline_step1}, is essential to provide dependable cues that reflect the user's actual intentions.

\subsection{Intention-Oriented: Constructing Query Response}
Upon restoring the user's query intent, the ensuing task is to formulate a response that is congruent with this discerned intent. Considering the widespread adoption of gaze-facilitated VR in tourism~\cite{ir-tourist}, it is plausible to envision IR systems developed within \name paradigm could be effectively applied for such use cases that requisite professional knowledge. Moreover, a small segment of users has expressed their anticipation for \name to incorporate various features inspired by other well-established systems, as shown in Section~\ref{sec:expectation}. In response to these insights, we advocate for a bifurcated approach in guiding system design, which encompasses an open design space for mounting such desired functionalities.

\subsubsection{Tools-Driven Knowledge Retrieval.}
While addressing the professional use cases mentioned above necessitates specialized knowledge domains, everyday scenarios also demand specific expertise. For instance, in a shopping context, comparing prices to online products requires access to an e-commerce database. Thus, we propose that a mature system within the \name paradigm should incorporate a module of diverse functionalities and a corresponding dispatcher. A particularly notable function is short-term memory, which, when integrated with \name's unique gaze modality, distinguishes it from similar functions in other paradigms. The interaction between gaze data and this short-term memory module—specifically, the criteria for triggering data recording and the format of such records—presents a complex and independent research topic, as well as the privacy issue raised by its always-on property.

\subsubsection{Intention-Guided Information Integration} Upon dispatching specialized functionalities, a critical subsequent step involves refining the output through a process of filtering and rephrasing, tailored to align with the user's specific query intentions. 

In the development of our comprehensive design framework, we adhered to two fundamental principles that consistently guided every aspect of it. The first principle is a steadfast focus on the user's intention. This user-centric approach ensures that every component of the system is designed and optimized to cater to the user's specific needs and goals. The second principle is the utilization of gaze tracking as a primary tool for uncovering user's query intentions, which is a unique feature introduced by \name paradigm. By leveraging gaze data, the system gains a nuanced understanding of what the user is focused on, thereby providing insights into their underlying intentions and interests. These principles are not just theoretical guidelines but are deeply integrated into the framework's architecture, influencing everything from understanding questions to answering questions.

\section{A preliminary implementation} \label{sec:implementation}

The aforementioned analysis based on a user-enactment study (Section~\ref{sec:findings}) already provides abundant insights for a system design framework within the \name paradigm. However, further implementation of a functional system (\RIII{henceforth referred to as 
\system}) in a user study could offer additional understanding of user interactions with such a system. More importantly, an evaluation study is instrumental in proving the concept of integrating gaze tracking in the \name paradigm and the effectiveness of our proposed design framework.

\subsection{\RIII{\system} Implementation}

\RIII{\system}\footnote{Open source code at: https://github.com/Sky-Wang326/gvoila.git} collects user query data using Pupil Labs Invisible eye-tracking glasses and posts requests to a remote server equipped with NVIDIA V100 for processing responses. As shown in Figure~\ref{fig:implementation}, users can initiate queries by speaking to \RIII{\system}, with the audio being transcribed using the Whisper~\cite{radford2023robust}. The local gaze-based temporal localization process computes key video frames, which, along with the query and gaze data, are sent to the remote server. Responses, comprising reconstructed queries, direct answers, and inferential thoughts, are returned and presented on a computer screen for users to evaluate.

We use SAM~\cite{kirillov2023segment} to segment users' interest objects at various scales according to gaze points. Caption generation is handled by KosMos-2~\cite{peng2023kosmos2}, while OWL-ViT~\cite{minderer2022owlvit} is used for object detection. KosMos-2 aids in depicting the overall context. Leveraging the advanced capabilities of large language models (LLMs) in creating natural language-based agents, we selected GPT-4 from the ChatGPT API for its exceptional comprehension, inference, and instruction-following abilities.\RIII{\system}'s prompt structure is based on the instruction-examples format, elucidating implicit inquiry form-factors (Figure~\ref{fig:exp1_category}) within the instruction part and further exemplified. \RIV{Prompts can be found in Appendix~\ref{appendix:prompt}, detailed implementation specifics can be found in Appendix~\ref{appendix:g-voila-implement}.}

\subsection{Baseline Implementation} \label{sec:baselines}

To reiterate, the distinctiveness of both the \name paradigm and \RIII{\system's implementation} stems from the incorporation of gaze information. Consequently, our primary comparisons involve systems that eliminate the influence of gaze. In \RIII{\system}, gaze has been instrumental in reconstructing user query intentions, specifically for temporally and spatially pinpointing users' interests. We implemented three baseline systems in an ablation manner, described as follows:

(1)\textbf{VOILA.} In this system, we eliminate gaze input. The gaze-driven spatial localization process is replaced with a confident object detection result, while the gaze-driven temporal localization process is substituted by randomly selecting unblurred video frames.

(2) \textbf{VOILA-T.} In this system, we retain the gaze-driven temporal localization process compared to VOILA. VOILA-T served as a baseline throughout our evaluation study.

(3) \textbf{VOILA-S.} In this system, we keep the gaze-driven spatial localization process compared to VOILA. 

\RII{We further considered two alternatives to the incorporation of gaze:}

\RII{(4)\textbf{VOILA-center.} In this system, we consider user's facing direction as their interest indicator, which is the center of camera's visual field.}

\RII{(5)\textbf{VOILA-saliency.} Considering the prevalent of saliency prediction techniques, where user's gaze trace on an image can be predicted by a deep learning model, we incorporated a saliency model to substitute gaze data. Implementation details can be found in the Appendix~\ref{appendix:saliency}.}

\section{Evaluation Study}
With the preliminary implementation of \RIII{\system}, we designed an evaluation study to prove the effectiveness of integrating gaze and the guideline of such integration proposed in the design framework.

\subsection{Evaluation Study Design}

\subsubsection{Scenarios and Participants.} \label{sec:exp2_participants}
The user study involves two daily-life scenarios: supermarket shopping and domestic living. We excluded the museum visiting scenario, previously considered in the user-enactment study,  due to the challenges in integrating a specialized database, which we were not able to acquire. As described in Table~\ref{tab:task_descriptions}, six tasks for each scenario were predefined, drawing upon the queries identified during the formative study. We designed these tasks with a twofold purpose: (1) To encourage participants to ask questions requiring varied applications of gaze tracking and egocentric views, as exemplified in Figure~\ref{fig:exp1_anticipation}. For instance, the ``Comparison'' task prompts participants to engage with multiple items simultaneously. (2) To offer a structured guideline for participants, assisting them in formulating queries, particularly in instances where they might struggle to think of questions.

We included 16 participants (8 for each scene \RIII{and have no overlap with previous enactment study}), comprising 8 males and 8 females, aged from 20 to 30 (with a standard deviation of 2.06). Each participant required approximately 240 minutes to complete the study, and they received a compensation of 30 USD per hour for their participation.

\begin{table}[]
    \centering
    \resizebox{\linewidth}{!}{
    \begin{tabular}{cp{5cm}|cp{5cm}}
    \toprule
        \multicolumn{2}{c}{\textbf{Supermarket shopping}} & \multicolumn{2}{c}{\textbf{Domestic living}} \\
         Task & Description & Task & Description \\
    \midrule
    Comparison & Query for details to make a choice among several products & Appliance Malfunction & Seek solutions to rectify the malfunction \\
    Completing Recipe & Figure out how to incorporate an ingredient into a cherished dish & Activity \& Health & Get advice for certain health concerns or activities \\
    Recommend & Dishes with ingredients in front of you & Snack \& Fruits & Choose from snacks and fruits for a specific purpose \\
    Knowledge & Query for a specific attribute of a product & Dressing Advice & Pick out the right outfit for tomorrow's plan \\
    Decision Making & Request \RIII{\system} to decide whether to buy a product based on certain criteria & Entertainment & Recommend recreational activities at home \\
    Strengthen Decision & Request \RIII{\system} to strengthen your resolve to resist a certain temptation & Small Talk & Suddenly desire intriguing information about a specific object \\
    \bottomrule
    \end{tabular}
    }
    \caption{Guiding Tasks for Evaluation Experiment}
    \label{tab:task_descriptions}
\end{table}

\subsubsection{Procedure.} \label{sec:exp2_procedure}
\begin{figure}
    \centering
    \subcaptionbox{\RIII{Participants' querying paradigm and systems used to generate responses.}\label{fig:exp2-4rounds}}
    {\includegraphics[width=0.7\linewidth]{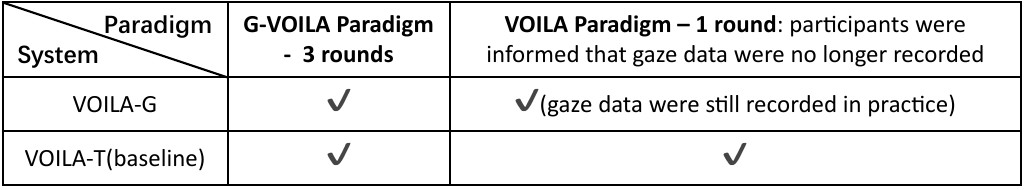}}
    \subcaptionbox{\RIII{Evaluation pipeline for each round of usage. Repeated for 4 rounds.}\label{fig:exp2-rounds-pipeline}}
    {\includegraphics[width=0.9\linewidth]{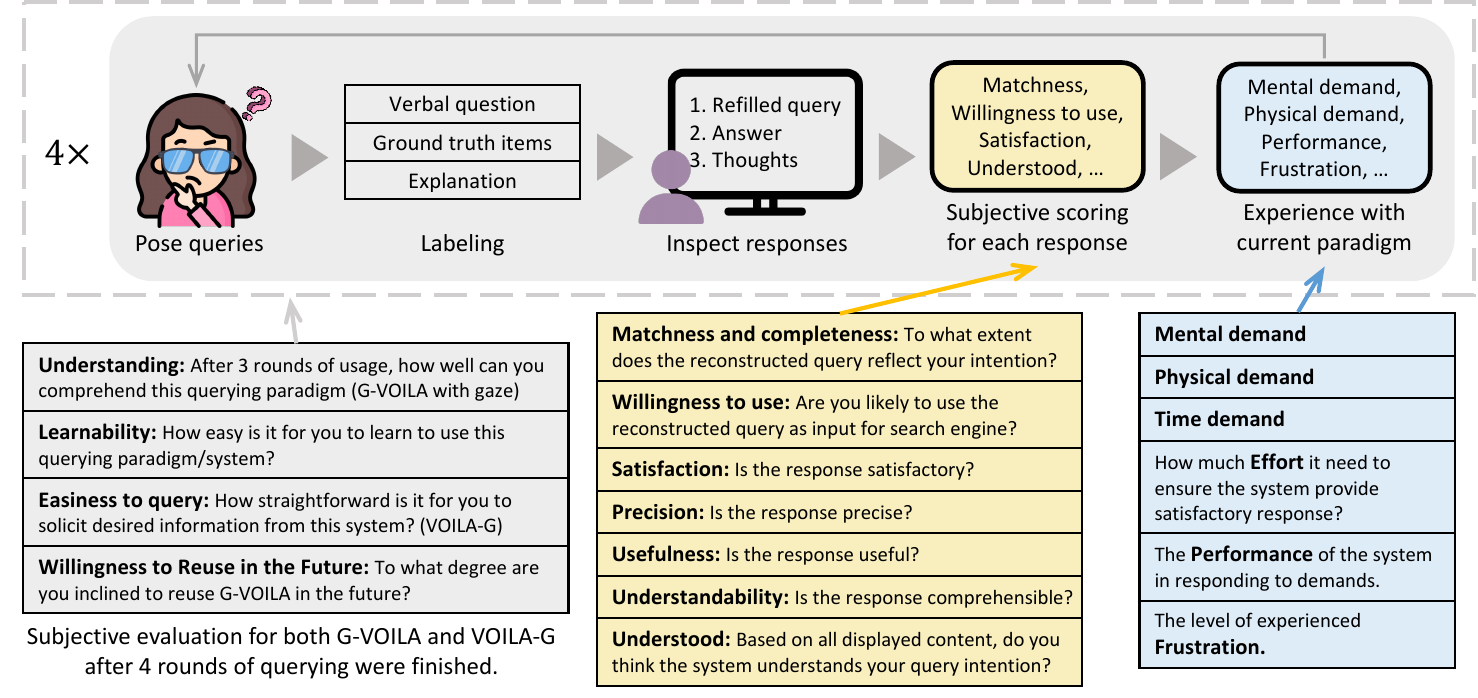}}
    \caption{\RIII{Procedure of evaluation study.}}
    \label{fig:exp2-design}
\end{figure}

Similar to the user-enactment study, the evaluation study comprises three sessions: an introduction session, a querying session in which participants engage in four rounds of using and evaluating, and a subsequent interview session. The introduction session follows the same format as the user-enactment study. We emphasized \RIII{\system}'s uniqueness of gaze tracking compared to other querying paradigms and encouraged participants to actively try out various phrasing strategies that can allow them to express themselves in a more convenient manner.

In the querying session, each participant takes part in four rounds of querying, as depicted in Figure~\ref{fig:exp2-4rounds}. Within the continuous three-round usage of the \name paradigm, participants are made aware of \name's ability to capture eye movements and are encouraged to experiment with different strategies for phrasing their questions. Simultaneously, participants are required to utilize the VOILA paradigm for one round. The VOILA paradigm is a modified version of the \name paradigm, with the gaze modality eliminated. Therefore, during the one-round usage of the VOILA paradigm, participants are informed that gaze data are no longer recorded; however, gaze data are still collected to support the response process of \RIII{\system}. The order of different paradigms is counterbalanced among participants.

The evaluation pipeline for each round is illustrated in Figure~\ref{fig:exp2-rounds-pipeline}. At the beginning, participants are asked to pose one query for each predetermined task. Subsequently, they are instructed to identify the objects they are inquiring about, which are later employed as ground truth to objectively calculate response recall and precision. For each query, both \RIII{\system and VOILA-T(baseline)} provide responses consisting of (refilled query expression, answer, thought). After examining the responses from both systems, which are displayed anonymously in a randomized order, participants are requested to subjectively rate the responses from both systems, with the understanding that waiting time should be disregarded. Finally, participants are required to rate their experience of querying within the current paradigm. \RIII{After the inspecting and scoring were done for current round, participants can move on to the next round of querying. Such ``querying, labeling, checking answer, scoring'' process will be repeated for 4 rounds. With such repetition, participants may become more familiar with \name paradigm and \system system.}

In the interview section, participants were asked to respond to the same questions posed in the subjective evaluation of the querying paradigm, along with some additional inquiries as outlined \Concise{in Figure~\ref{fig:exp2-rounds-pipeline}'s list (bottom-left).}

\subsubsection{Evaluation Metrics.}
For the objective evaluation, we employed recall score and precision score. The ground truth objects were identified by participants in each round of usage. We recognized items emphasized in the responses as predicted values. Given that the predicted values are in open-vocabulary, we cannot calculate the objective accuracy score.

Participants rated the subjective metrics on a 7-point Likert scale. For each query response, the metrics are \Concise{outlined in Figure~\ref{fig:exp2-rounds-pipeline}'s list (bottom-middle).}

For the experience with the querying paradigm, the metrics are \Concise{outlined in Figure~\ref{fig:exp2-rounds-pipeline}'s list (bottom-right).}

\subsection{Results and Analysis} \label{sec:exp2_findings}

\begin{table}[]
    \centering
    % \resizebox{\linewidth}{!}{
    \begin{tabular}{cccccccc}
    \toprule
         & & \system & VOILA-S & VOILA-T & VOILA & \RII{VOILA-center} & \RII{VOILA-saliency} \\ 
    \midrule
    Recall & \textbf{All*} & 89.1\% & 83.84\% & 79.25\% & 76.91\% & 78.44\% & 78.76\% \\
     & Explicit & 97.28\% & 94.12\% & 94.12\% & 93.6\% & 92.81\% & 93.86\% \\
     & \textbf{Ambiguous*} & 80.11\% & 72.54\% & 62.91\% & 58.57\% & 62.67\% & 62.19\% \\
    Precision & \textbf{All*} & 83.83\% & 78.4\% & 69.83\% & 67.62\% & 71.8\% & 72.86\% \\
     & Explicit & 95.79\% & 92.73\% & 93.44\% & 91.23\% & 91.6\% & 93.03\% \\
     & \textbf{Ambiguous*} & 70.7\% & 62.65\% & 43.9\% & 41.7\% & 50.06\% & 50.7\% \\
    \bottomrule
    \end{tabular}
    
    \caption{\RII{Objective Response Evaluation. Recall rate and precision score for both \RIII{\system} and the baseline models. Queries related to specific objects in the scenario are classified into explicit and ambiguous categories based on whether the user mentions the target object's name in their query expression. The Wilcoxon signed-rank test was employed for all comparative scoring, with significant differences denoted by \textbf{bold*} labels. We discussed the results of VOILA-saliency in Appendix~\ref{appendix:saliency}.}}
    \label{tab:exp2_result}
\end{table}

\subsubsection{Objective Response Score} \label{sec:exp2_objective_score}

With labeled querying objects for each question, we calculated the recall rate and precision for \RIII{\system} and the baseline. We examined whether the key objects in the ground truth label appeared in the response and identified semantically parallel objects as detected keywords. We excluded queries that were not related to specific content in the scenario, then categorized the remainder into \textbf{Explicit Query} and \textbf{Ambiguous Query}based on whether key object names were spoken in the query. Table~\ref{tab:exp2_result} presents the average recall and precision for each category.  \RIII{\system} statistically outperforms VOILA in ambiguous queries(recall: $Z=184.5$, $p<.001$; precision: $Z=497.5$, $p<.001$) and all queries(recall: $Z=295.5$, $p<.001$; precision: $Z=873.0$, $p<.001$), while slightly surpassing VOILA in explicit queries(recall: $Z=14.0$, $p=.013$; precision: $Z=49.0$, $p=.002$).

\RIII{Our ablation study indicates that using gaze as a spatial cue substantially improves \RIII{\system}'s performance. Notably, when leveraging gaze solely as a spatial indicator, incorporating its temporal properties further refines \RIII{\system}'s ability to comprehend ambiguous queries (from 72.5\% to 80.1\%). An improvement in both recall and precision score 
(\RIII{\system} vs. VOILA-T, VOILA-S vs. VOILA) demonstrates the efficacy of integrating gaze's spatial localization process. However, (\RIII{\system} vs. VOILA-S)exhibits a more substantial performance growth than (VOILA-T vs. VOILA),  which suggests that the advantage conferred by incorporating gaze's temporal localization process is more pronounced when gaze serves as a spatial indicator.}

\RIII{The precision improvement (29\%) with \RIII{\system} is notably higher than recall enhancement (22\%), especially for ambiguous queries. This suggests that \RIII{\system} effectively leverages gaze data to discern user interest among multiple potential query-related items in the visual field.} For example, the implemented systems may occasionally include extra items they presume to be of users' interest for explicit questions. Furthermore, when dealing with highly ambiguous queries, such as simply stating ``calorie content'' without using restrictive pronouns like ``which one between these two'', baseline systems may enumerate the calorie content for all detected food in the visual content. However, \RIII{\system} can pinpoint objects of interest to reduce the occurrence of both issues.

\RII{VOILA-center shows enhanced performance compared to VOILA, indicates the correlation between eye and head movements, with the latter serving as a reasonable proxy for inferring user intent. Nevertheless, head direction is a less precise indicator of intent compared to gaze (VOILA-center vs. VOILA-S). See Section~\ref{appendix:saliency} for a detailed discussion on VOILA-saliency results.}

\takeaways{
\RIV{\textit{Take-aways:} Incorporating gaze as a temporal and spatial indicator can both enhance the ability to discern user's interest. A synergistic catalytic effect exists when incorporating both temporal and spatial properties. Head direction can be used as a less precise indicator of user's intent. Existing saliency models fall short for \name's use case (see Appendix~\ref{appendix:saliency}).}
}

\begin{figure}
    \centering
    \subcaptionbox{Subjective scoring for queries\label{fig:exp2-query-subjective}}
    {\includegraphics[width=0.6\linewidth]{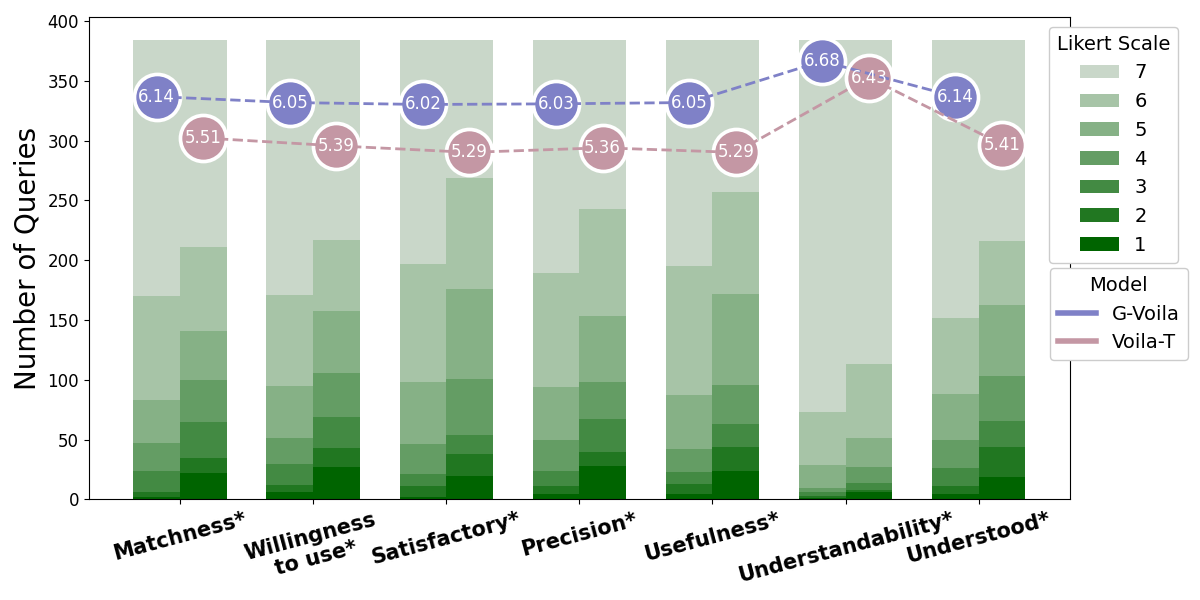}}
    \subcaptionbox{Subjective scoring for strategies\label{fig:exp2-round-subjective}}
    {\includegraphics[width=0.39\linewidth]{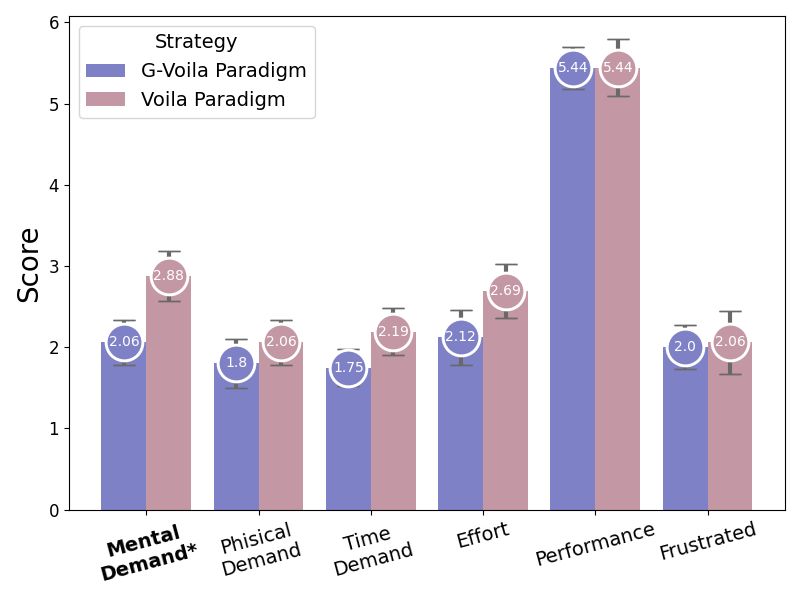}}
    \caption{Subjective Scoring Analysis. The Wilcoxon signed-rank test was utilized for all comparative scoring, with significant differences indicated by \textbf{bold*} labels. (a) A stacked bar plot representing each query's score, where a higher value signifies better performance. (b) A bar plot comparing gaze and no-gaze querying strategies, with lower values indicating superior results (excluding the performance bar).}
    \label{fig:exp2-result}
\end{figure}

\subsubsection{Subjective Score for Queries.}
\RIII{The subjective score for \RIII{\system} and VOILA-T, as presented in Figure~\ref{fig:exp2-query-subjective}, indicate a statistically significant preference for \RIII{\system} via a Wilcoxon signed-rank test across all metrics. Participants favored \RIII{\system}'s query reformulations, citing a higher concordance with their intended meaning and a stronger propensity to use these queries as text-based search input. \RIII{\system} was also rated higher in satisfaction, utility, and perceived accuracy over the baseline. However, the understandability of \RIII{\system}'s outputs showed only a slight increase, as LLMs are already adept at producing intelligible text, which gaze data can contribute minimally to. Overall, when evaluating answers and thoughts collectively, participants felt that \RIII{\system} more effectively captured their query intentions with the aid of gaze data.}

\subsubsection{Subjective Score for Paradigms.}

To ensure impartial evaluation, the order of response presentation from both models was randomized and anonymized. Participants rated the used paradigm for each round based on their usage experience during querying and displayed responses. We selected the score for the final round of applying \name paradigm compared to the only round of applying VOILA paradigm, as shown in Figure~\ref{fig:exp2-round-subjective}. Participants' preferences for the two strategies varied individually, with \name holding an overall advantage. Subsequently, we conducted interviews with each participant, and the results thereof were discussed in Section~\ref{sec:interview-part1}.

\begin{figure}
    \centering
    \subcaptionbox{Ambiguous count and query score\label{fig:exp2-round-query}}
    {\includegraphics[width=0.49\linewidth]{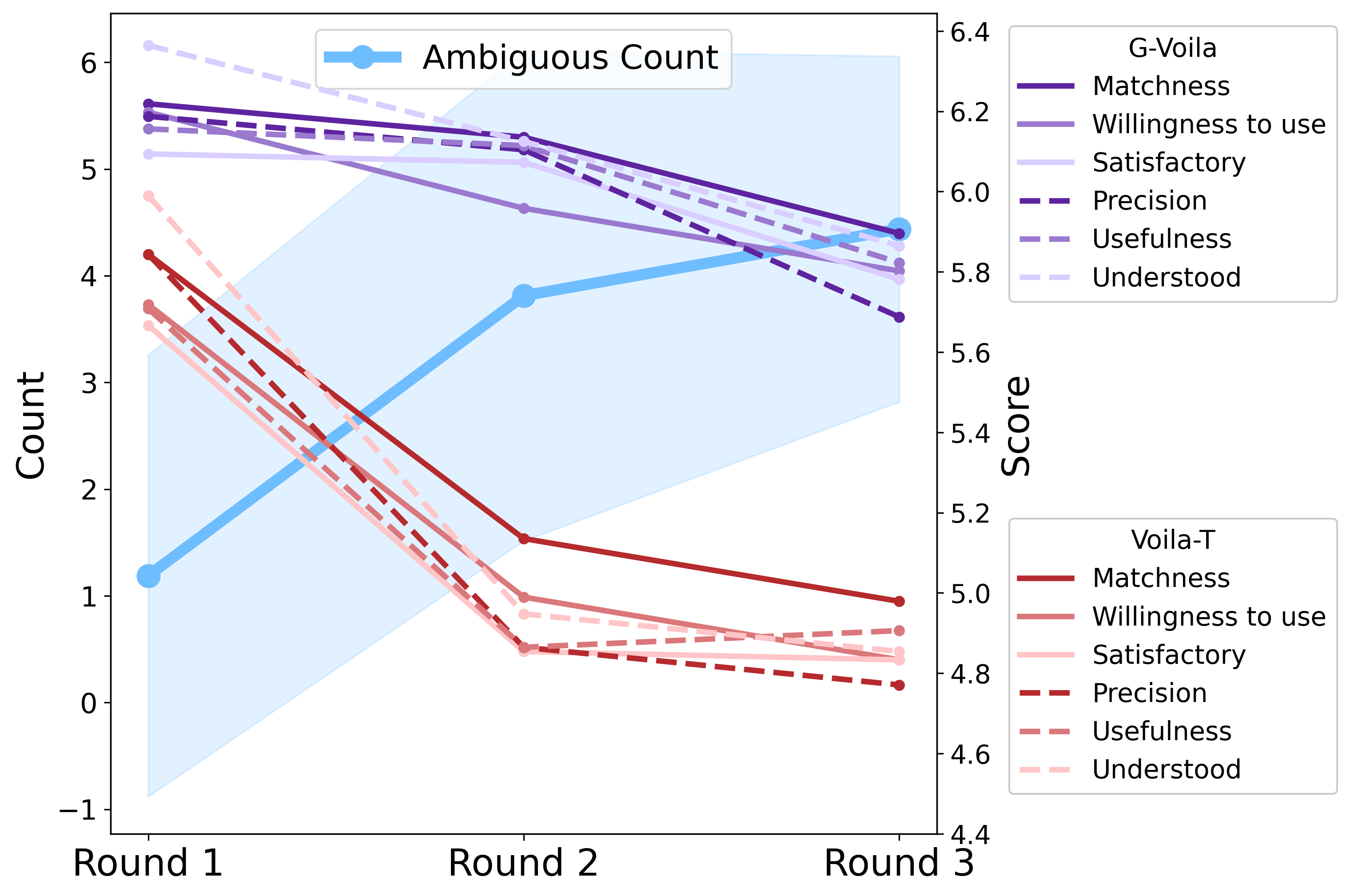}}
    \subcaptionbox{Ambiguous count and round score\label{fig:exp2-round-round}}
    {\includegraphics[width=0.49\linewidth]{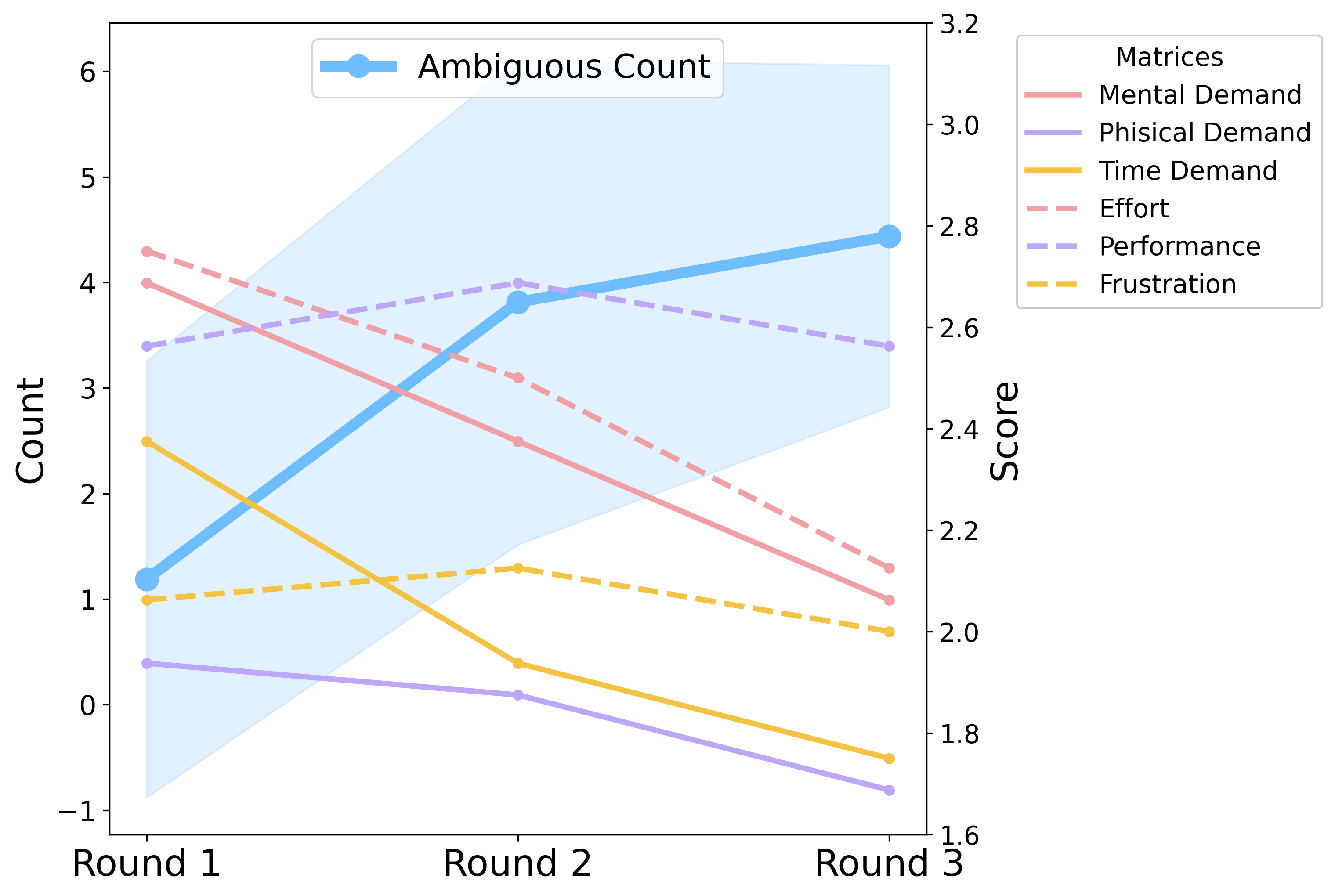}}
    \caption{User's scoring changes of three-round usage of \name{}. Both graphs show ambiguous query count for each round. (a) illustrates how 6 questions' average scoring changes for each evaluation matrix. (b) illustrates scoring change for \name{}'s usage. The ``performance'' matrix is subtracted by 8 to better fit in the plot.}
    \label{fig:exp2-round-anal}
\end{figure}

\subsubsection{Three-round Usage of \name paradigm}
Figure~\ref{fig:exp2-round-anal} shows the trajectory of participant ratings over three rounds usage of \name{} paradigm. Following each round, our researchers communicated with the participants to facilitate a better understanding of \name{}'s functionality and encouraged them to express queries more naturally, in accordance with their daily habits. 

Figure~\ref{fig:exp2-round-query} displays the ambiguous question count and the average query scores for both models across the three rounds. We omit understandability from the analysis due to its insignificance. A trend emerges where participants, growing accustomed to \name{}, posed their queries with increased ambiguity. The heightened ambiguity had a minimal impact on \RIII{\system}'s performance compared to baseline, benefiting from gaze-derived interest information.

Figure~\ref{fig:exp2-round-round} illustrates the ambiguous question count and scores for each round's overall experience with \name{}. It should be noted that the original rating for performance was higher-is-better; however, to better fit the plot, we modified the performance as $8-x$. As participants became more familiar with \name{} and used more arbitrary expressions, they tended to rate \name{} higher in all aspects. This finding aligns with the learning process typically observed with the adoption of any convenient new technology.

\takeaways{
\RIV{\textit{Take-aways:} As users became more familiar with \name{}, they increasingly leveraged its capabilities by posing more ambiguous queries, which have lower performance degradation on \system and better user experience.}
}

\subsection{Feedback and Interviews}

In the interview, participants were asked about their rating criteria and thoughts on each metric when evaluating \name usage. They also provided overall user experience scores and explanations.

\subsubsection{Rating for Each Round.} \label{sec:interview-part1}

Participants were interviewed about their experiences with each round usage of \name, covering cognitive, physical, and time demands, effort, performance, and frustration.

Most participants regarded effort as a combination of cognitive, physical and time demands, which was reduced as they adapted to the querying style during the experiment (or in potential long-term use) and formed a habit. Compared to the traditional text-based query approach, participants often cited two advantages: (1) simplified queries due to casual conversational and omitted information, and (2) on-demand querying without extra hand operations, such as taking out a device or operating software. These factors directly led to lower demands. However, P14 noted that, compared to search engines that provide multiple options, \name and chatbot systems offer more targeted answers, which may increase cognitive pressure to ask questions precisely. Meanwhile, P10, P11, and P14 found that controlling their eye movements physically demanding, while others regarded gaze as a natural behavior.

Similarly, performance is considered a combination of effort and response accuracy, while frustration is related to expectation and misidentification. The response accuracy of \RIII{\system} dropped slightly when queries became more ambiguous and participants adjusted their expectations as well. P7 and P13 stated that they were expecting delightful surprises brought by unexpected associations rather than overly focusing on context information for those questions where \RIII{\system} successfully identified their fixation content. Furthermore, P15 mentioned that \textbf{\name}'s attempts to combine context and provide more comprehensive answers could alleviate frustration when misidentification occurred.

\begin{figure}
    \centering
    \includegraphics[width=0.4\linewidth]{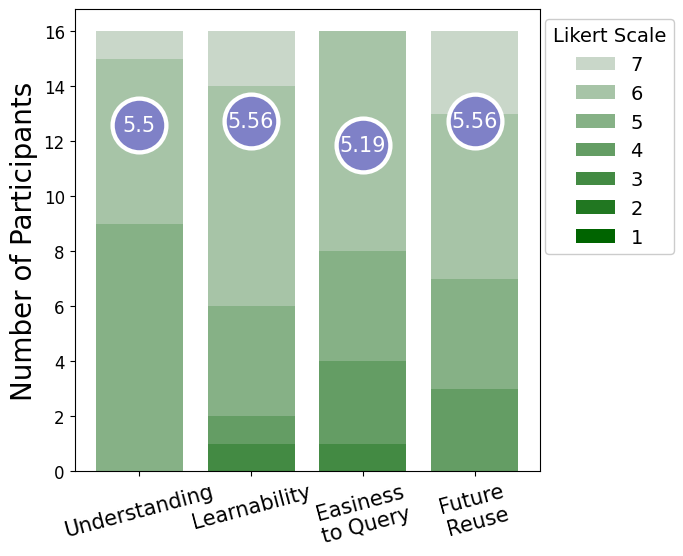}
    \caption{Overall Evaluation. A stacked bar plot illustrating participants' scores on a Likert scale regarding their overall experience with \RIII{\name paradigm and \system}, as reported in the follow-up interview. In this plot, a higher value signifies a better result.}
    \label{fig:exp2-voila-score}
\end{figure}

\subsubsection{Rating for \RIII{\name paradigm and \system}}
As shown in Figure~\ref{fig:exp2-voila-score}, participants exhibited a positive preference for \RIII{\system}.
The majority of participants reported that after 2-3 rounds of use, they were able to better understand \RIII{\system}'s capabilities and limitations, as well as master the collaboration between eye movements and query expression. Their inquiry style became more aligned with colloquial and daily expressions, which differs significantly from that of commonly used search engines, making it relatively easy to form a question. P3 and P10 mentioned that bearing the risk of incorrectness might increase the difficulty of asking questions. P14 noted that information is substantially fully presented to \RIII{\system} in either natural language or gaze behavior, providing a choice for output channels.

Due to \name's on-demand nature and its ability to alleviate the burden of daily questioning, most participants expressed a high willingness to reuse it in the future, yet they had certain expectations for its development. P16 mentioned that \RIII{\system}'s advantage lies in its ability to comprehend spatial knowledge, and he hoped that this capability could reach the same level as ChatGPT's text comprehension ability. Furthermore, P3 expressed concerns about the popularity of head-mounted devices, as \RIII{\system} relies on such a platform. Other suggestions beyond gaze and vision were given, including the creation of user profiles or integration with smart home control systems.

\section{Discussion and Future Work} 
\label{sec:discussion}

In this section, we discuss \name{}'s major results, findings, and limitations. We also provide insights for future work.

\subsection{\RIV{A Summary and Discussion of Findings in User-Enactment Study}} \label{sec:discussion_findings}

\subsubsection{Ambiguity: Reasons for users omitting certain information in their query expressions}
\RIV{The taxonomy introduced in Section~\ref{sec:findings} indicates various categories of ambiguity that may arise during the query formulation process with \name. Statistical results, as depicted in Figure~\ref{fig:exp1_catestat}, reveals a significant incidence of ambiguous queries, with a notably high frequency of pronoun usage exceeding our initial expectation. We further discuss several underlying factors that this tendency to generate queries lacking explicit environmental context and precise intent may stem from.}

\begin{enumerate}
    \item \textbf{It is time-consuming or cognitively demanding to express certain information.} When users are unfamiliar with the precise name of an item, they tend to use pronouns and object categories as a way to describe it. For instance, in a museum setting, they might refer to a ceramic piece as ``this porcelain'' instead of reading the name on the exhibit label. Similarly, when referring to multiple items, users prefer to use pronouns to indicate the range of objects they are comparing. For example, in a supermarket, they might say ``these few'' when comparing different types of fish rather than listing specific names.

    \item \textbf{Users struggle to create well-organized expressions for their inquiry intent.} In some cases, users may not know what the target object is and can only use pronouns to refer to it. A kitchen novice wanting to know how to eat red amaranth might describe it as a spinach-like vegetable with purple meridians on the leaf, without knowing the proper name. Additionally, users may associate the inquiry target with existing knowledge, leading to the use of incorrect or inaccurate vocabulary in their descriptions. For instance, when asking how a ``relief'' is made on a porcelain plate in a museum, it is confusing because traditionally understood relief is on architects or wooden artifacts.

    \item \textbf{Users become immersed in a situation and neglect to describe certain constraint information.} When looking for activities at home, users may overlook keywords like ``indoor'' but still expect to receive suggestions suitable for their situation, such as not wanting outdoor activities like skiing or swimming. In another scenario, users may not be aware that their current situation imposes constraints on the information retrieval process. An ESL user in a museum, for example, might select the most appropriate definition of an unfamiliar English word on an exhibit label based on the context when using a dictionary without realizing the importance of the context in their search.
\end{enumerate}

\subsubsection{\RIV{Eye-movement pattern throughout each querying and it's coordination with mouth.}}

\RIV{Through rigorous quantitative analysis, we have delineated several patterns that are congruent with established gaze-related behaviors during task execution, as well as novel patterns specific to \name. Users tend to focus more on the relevant objects than irrelevant ones when conveying queries, whereas the distribution of this relevancy does not remain uniform throughout the querying process. Analogous to the phenomenon where ``gaze often moves on before the last act is complete''~\cite{land2006eye-movements-and-actions}, users initially direct their gaze toward objects that are pertinent to the content of their query. As the query progresses, we observe a greater deviation in gaze towards the end.}

\RIV{Conversely, our findings indicate that users typically shift their attention to items of interest prior to initiating their queries, with a pre-emptive focus averaging 4.62 seconds. Most notably, a distinctive gaze pattern associated with the querying task is the existence of eye-mouth coordination, where users are most probably looking at their intended objects before the pronouns were spoken.}

\RIV{Beyond the general quantitative analysis, our researchers have noted a "turning back" behavior among most participants when annotating data. This behavior is characterized by users redirecting their gaze to inquire about objects they had previously disregarded. Additionally, variations were observed in participants' behaviors at the conclusion of their queries. While the majority of participants remain stationary until the entire question is articulated, a subset exhibits a tendency to move away promptly after formulating their question.}

\subsection{\RIV{Response Precision and User Experience}} \label{sec:discussion_lowacc}

\RIV{The precision and recall metrics for \system when addressing ambiguous queries, as depicted in Figure~\ref{fig:exp2-objective-score}, do not achieve a production level. This shortfall is primarily due to the reliance on open-source models for constructing the visual system of \system without much engineering effort. The limitations are twofold: (i) the performance of the visual understanding models does not meet the product standards for real-world applications, and (ii) there is a discrepancy between the distribution of the training dataset (internet-sourced images) and our operational environment (snapshots from egocentric videos). Notably, approximately half of the study participants (n=7) expressed potential frustration with instances where \system erroneously identified objects. This necessitates a discussion on the potential impact of suboptimal performance on user experience and the interpretation of experimental results.}

\RIV{However, \system has proven its conceptual validity by successfully integrating gaze tracking, as evidenced by both objective and subjective evaluations presented in Table~\ref{tab:exp2_result} and Figure~\ref{fig:exp2-query-subjective}. Participants rated \system more favorably on all subjective measures compared to the VOILA-T baseline with statistically significant differences. This suggests that even with room for performance enhancement, \system effectively serves as a proof-of-concept prototype. }

\subsubsection{\RIV{User Preferences and Trust in Response Precision}}

\RIV{Participants showcased divergent preferences concerning the precision of responses to similar questions. When inquirying about objects, \system sometimes provide answers that exceeded the scope of the user's intention, particularly for ambiguous queries. Some users had higher expectations for a precise response, while others demonstrated higher tolerance for redundant replies, as long as the target information was prominent and could be quickly located. For example, Participant 13 states his/her preference for more divergent answers, rather than being limited to the specific question and visual content.}

\RIV{One possible factor revealed in our afterwards interview is that individual's information retrieval habits may influence their expectations for precision. Participant 11, accustomed to converstional interface like ChatGPT, paid more attention in phrasing queriesto elicit precise responses. Conversely, Participant 15, who predominantly uses search engines, feels satisfied with a broader explanations provided by \system when it attempts to encompass his/her intention more fully.}

\RIV{Moreover, the expression style users presented in their final round of querying was related not only to their personal linguistic style but also to their trust in the system's capabilities. Some participants abandoned the use of complete sentences and resorted to terse phrases such as ``vitamin content'', ``want to exercise abs'', and ``eyes feel dry''. In subsequent interviews, these participants had high praise for \system's eye-tracking and visual abilities, indicating a correlation between their trust in the system and their expression style.}

\subsection{Future Deployment Potential}
\system relies on large-scale computer vision and language models, and as a result, is subject to their limitations. Firstly, although large language models have demonstrated impressive understanding capabilities, the text generation speed of the API is the primary factor contributing to waiting times, as mentioned by participants in their feedback. Secondly, the understanding capabilities of computer vision models have not yet reached a high level, which limits the accuracy of our system in recognizing objects that users gaze at. Moreover, there is still a discrepancy between the training datasets of currently available large-scale vision models and real-world ego-centric data, such as the recognition of fruits in a supermarket being significantly affected when wrapped in plastic film. Furthermore, to provide services in high IR demand scenarios, such as museums, integration with expert knowledge is necessary. Last but not least, the modules employed in this article to handle gaze, vision, and text modalities are relatively independent from each other, resulting in insufficient accuracy in spatially and temporally locating objects of interest to users and occasionally leading to deviations in understanding user intent. To better integrate these data, a multi-modal large model with excellent performance is needed.

\subsection{Beyond Gaze and Voice}
\name currently relies on gaze and voice input for query-based information retrieval, whereas we have discovered other data modalities that might help for natural querying expression. One typical data modality is gesture, which has been widely used in Human-Computer Interactions and can also serve a pointing purpose. Activity data captured by the inertial motion unit (IMU) may indicate the user's movements and head motion, such as stopping at some place, may indicate an increase of interest and can also help to stabilize gaze points. Moreover, facial expressions can also reveal the user's attitude towards currently looking content. We expect to add more insightful data modalities to \name{} in the future.

\section{conclusion} \label{sec:conclusion}
To conclude, we first envisioned a future information querying paradigm, namely \name, which combines user's gaze data, visual field and voice-based natural language queried. We revealed users inherent ambiguous and colloquial phrasing manner when querying with the \name paradigm in a user-enactment study. Based on quantitative analysis of collected data, we discovered a mouth-eye coordination in users' expression behavior. Inspired by this study, we proposed a design framework for \name which addresses the key aspect of gaze incorporation. We implemented a proof-of-concept \name assistant by harnessing the advanced deep learning technique and employed it for a controlled user study. We demonstrated the effectiveness of \name by achieving 89\% recall rate and 84\% precision score, surpassing the baseline method in both objective and subjective scoring metrics.

\begin{acks}
This work is supported by the Natural Science Foundation of China (NSFC) under Grant No. 62132010, Young Elite Scientists Sponsorship Program by CAST under Grant No.2021QNRC001, Tsinghua University Initiative Scientific Research Program, Beijing Key Lab of Networked Multimedia, Institute for Artificial Intelligence, Tsinghua University, Undergraduate / Graduate Education Innovation Grants, Tsinghua University, and Beijing National Research Center for Information Science and Technology (BNRist).
\end{acks}

\bibliographystyle{ACM-Reference-Format}
\bibliography{sample-base}

\appendix
\section{\system Prompt} \label{appendix:prompt}
The prompt used for \system is shown in Table~\ref{tab:prompt}.
\begin{table}[htbp]
    \centering
    \resizebox{\linewidth}{!}{
    \begin{tabular}{l p{14cm}}
    \toprule
    \ding{228} Overall template: & 
{\scriptsize VOILA is designed to be able to assist with visual question answering task. With the input of a series of ego-centric snapshot description and user's query question, VOILA can (1) answer the question directly and (2) provide query input for web search engine. VOILA is able to understand large amounts of images and videos, then answer user's question with its knowledge and the help of search engine. VOILA can not directly read images or videos, but a list of visual understanding tools will provide textual information about the visual content, based on which VOILA can infer knowledge about the visual content. The textual information provided may be vague and general, but VOILA should do its best to organize the information to understand the visual content. 
The whole visual assitant is able to capture ego-center snapshots and user's eyegaze attention coordinate, based on which several visual understanding tools will generate textual information. When the user asks questions knowing that VOILA has already understood the visual content and his/her eyeygaze interest, so his/her question may be vague by using pronoun or skipping intent words. VOILA should do its best to infer user's query intent by fulfilling the missing information. With the fulfilled query content, VOILA should (1) answer the questions directly and (2) generate proper and unambiguous query input for web search engine.
When inferring user's query intent, Voila should follow steps below: (1) figure out whether user's query contains ambiguous words, such as pronouns or words can not be understand depend solely on query text. (2) if yes, figure out the possible meaning of the ambiguous words based on the textual information provided by visual understanding tools and user's query question. (3) if no, check out whether the context or interest information might be related to user's query. If yes, combine it with query text properly. If no again, infer user's query intent solely based on the query text. 
Overall, VOILA is a powerful visual dialogue assistant tool that can integrate visual understanding content and user's ambiguous query to answer user's question and generate proper query input for web search engine.

------

VOILA takes the following information type as input: {{inputs}}

------

VOILA output a response in format (MUST BE IN JSON FORMAT):

```json
\{

"thought": string \textbackslash\textbackslash Explain the process of inferencing user's unambiguous query intent, use the textual information provided by visual understanding tools and user's query question

"answer": string \textbackslash\textbackslash The answer to user's question

"query": string \textbackslash\textbackslash The query input for web search engine

\} 
   }
\\

    \midrule
    \ding{228}~Input type & 
\ding{222}Context Caption: A textual description of the whole visual content, generated by the visual captioning tool.

\ding{222}~Interest Caption: A textual description of the user's eye gaze interest, generated by a visual captioning tool -- a list of textual descriptions of the object that the user is looking at. The recognition result might be inaccurate, and the input is the top 3 descriptions with the highest confidence.

\ding{222}~OCR: Extracted text from the whole vision field, generated by OCR tool, the recognition result can be considered highly inaccurate except for understandable phrases.

\ding{222}~User Query: The user's query question, may be vague by using pronouns or skipping intent words. Query text is a transcript using a speech recognition tool, which may be inaccurate if you find some words hard to understand.\\

    \bottomrule
    \end{tabular}
    }
    \caption{\textbf{Prompt Template of Voila-G} }
    \label{tab:prompt}
\end{table}

\section{\system Implementation Specifications} \label{appendix:g-voila-implement}

\RIV{\subsection{Device and Workflow}}

\RIV{We separated the devices as \textbf{[Local device]} and \textbf{[Remote device]}, which is more compatible with future deployment. Local devices are tasked with executing algorithms of lower computational complexity, such as data preprocessing, while the remote server is designated for algorithms that are more computationally intensive. An additional critical aspect of our setup is the optimization of data transfer between the Local device and the Remote device to facilitate real-time response capabilities in future iterations, which means transmitting entire video clips is impractical. Therefore, we temporally located at most three representative frames on the local devices and transmitted them to the server for further processing. }

\begin{figure}[h!]
    \centering
    \includegraphics[width=0.8\linewidth]{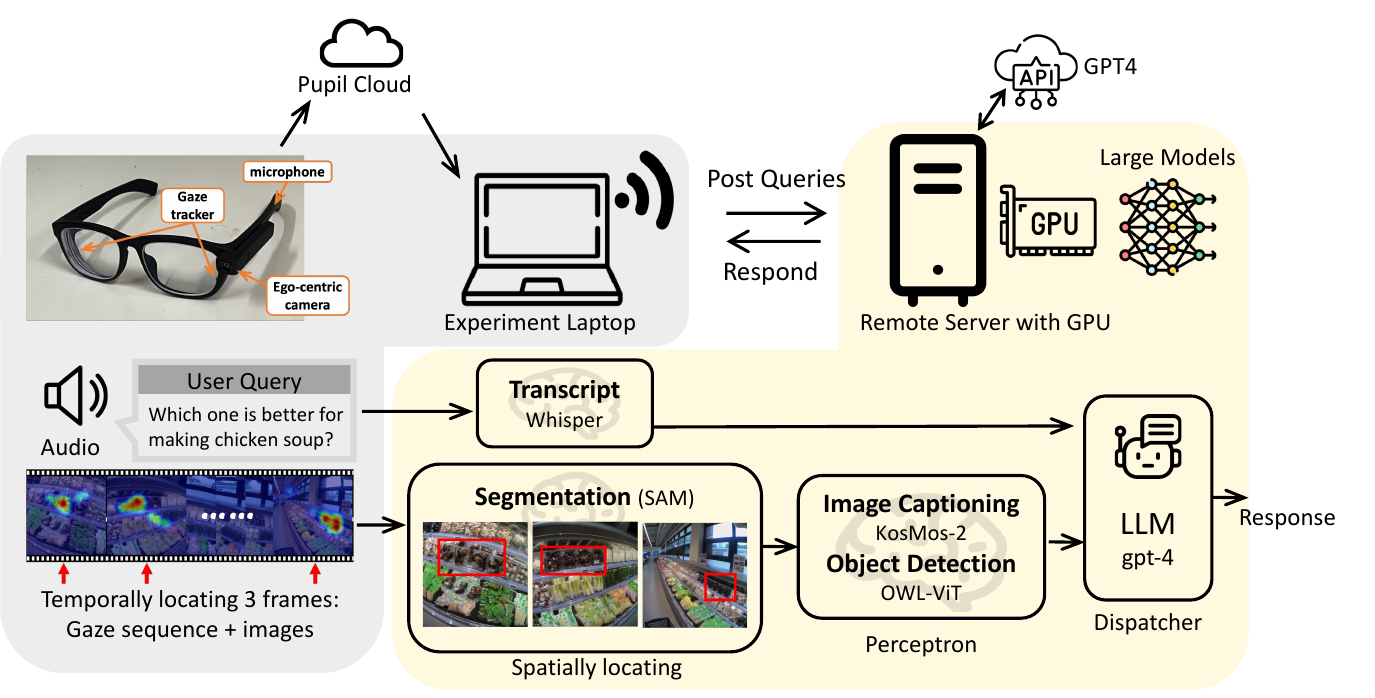}
    \caption{\RIV{A Preliminary Implementation: \RIII{\system's} pipeline.}}
    \label{fig:implementation}
\end{figure}

\RIV{\system's implementation is shown in Figure~\ref{fig:implementation}.}

\RIV{\begin{itemize}
    \item \textbf{[Local device]} The Pupil Lab Invisible glasses and a companion smartphone (OnePlus 8T O).
    \begin{itemize}
        \item Usage details: The Pupil Lab Invisible glasses require a connection to a companion smartphone to be driven, which is not depicted in this illustration figure. To reduce potential distractions, participants were advised to carry the smartphones in their pockets throughout the experiment.
        \item Data transfer: Upon completion of each round of queries, the collected data were uploaded to Pupil Cloud and subsequently downloaded to our experiment laptop.
    \end{itemize}
    \item \textbf{[Local device]} Experiment laptop - MacBook Pro with Intel processor and running the Ventura OS.
    \begin{itemize}
        \item Device requirements: our experimental setup is compatible with a broad range of laptops and OS.
        \item Usage and data transfer: We preprocess the data on the local devices and send the data to a remote server equipped with high-performance GPUs for further analysis.
        \item Response display: The final results returned by the remote server are presented to participants through a streamlined user interface, which is developed using REACT.
    \end{itemize}
    \item \textbf{[Remote device]} GPU server - either two NVIDIA V100 32GB GPUs or a single NVIDIA A100 80GB GPU.
    \begin{itemize}
        \item Device requirements: The total GPU memory requirement for \system is estimated to be around 23GB.
        \item Usage and data transfer: The server can handle the computational demands of multiple complex deep-learning models. After completing the generation of textual transcripts and descriptions for the submitted data, \system formulates a query utilizing the prompt outlined in Table~\ref{tab:prompt} for the GPT-4 API. The server then proceeds to cleanse the data format of GPT-4's output and relays the refined response back to the local laptop.
    \end{itemize}
\end{itemize}}

\RIV{\subsection{Data Preprocessing Workflow}}

% \textbf{Data preprocessing workflow:}

\RIV{\begin{itemize}
    \item A critical component of voice-interactive systems is the detection of the initiation and conclusion of user interactions, such as keyword spotting and push buttons. In our experimental framework, participants activate queries by directly speaking their questions. Thus, our researchers manually annotated the start and end points of each query using the audio editing software Audacity. For the prospect of real-time interaction—a feature not inherently supported by the Invisible Glasses due to the lack of an input mechanism—we integrated a WiFi-based button and other supportive (not inherently supported by the Invisible Glasses) for the prospect of future real-time interaction, not utilized in the current experiment to maintain simplicity.
    \item With the interval flag of each query, we programmed the local devices to selectively extract three frames for each query and crop out the associated gaze data sequences. The chosen images, gaze data, and the audio of the user's query were then posted to the server. The frame selection is guided by the criteria detailed in Section~\ref{sec:gaze_pattern}, where we prioritize high-resolution frames from fixation points that are (1) notably longer in duration and (2) located proximate to the query's initiation timestamp. We refrained from using pronouns as a cue for frame selection, as generating transcripts to identify such cues would necessitate the employment of Whisper and additional Natural Language Processing (NLP) models. Given their substantial computational demands, these models are impractical to run on local devices.
\end{itemize}}

\RIV{\subsection{Remote Algorithm Pipeline}}

% \textbf{Remote algorithm pipeline:}

\RIV{In the remote processing phase, we leverage deep learning models whose configurations are documented in Table~\ref{tab:imp_models}. The generation of context descriptions is provided by KosMos-2 ~\cite{peng2023kosmos2} wrapped up with gradio since Kosmos-2's requirements are complicated and more suitable to run in docker. Meanwhile, interest descriptions are managed by OWL-ViT~\cite{minderer2022owlvit}. We choose potential interest objects by (1) sorting bounding boxes based on the Intersection over Union (IoU) and (2) filtering them by a confidence score threshold. Due to the risk of misidentification in object detection tasks, our pipeline outputs three candidate labels for each object of interest, which increases the likelihood that the user's intended object is among the identified candidates (ensure recall).}

\begin{table}[]
    \centering
    % \resizebox{\linewidth}{!}{
    \begin{tabular}{c|cp{4cm}cc}
    \toprule
         & version & specified parameters & input data & output data \\
    \midrule 
    Whisper~\cite{radford2023robust} & large, multilingual, v2 & - & audio file & transcribed query text \\
    SAM~\cite{kirillov2023segment} & vit\_h(default) & - & image + gaze & b-box \\
    KosMos-2~\cite{peng2023kosmos2} & patch14-224, 1.6B params & description\_type=``Detail'' & image & caption \\
    OWL-ViT~\cite{minderer2022owlvit} & large, patch14 & - & image & (type, b-box, score) \\
    GPT-4~\cite{gpt4} & model0613, api0515 & max\_tokens=1500, \newline temperature=0, top\_p=1, \newline frequency\_penalty=0,  \newline presence\_penalty=0.6 & see prompt & see prompt \\
    \bottomrule
    \end{tabular}
    % }
    \caption{\RIV{Deployed deep learning models and API for \system. ``b-box'' is short for bounding box.}}
    \label{tab:imp_models}
\end{table}

\section{VOILA-saliency} \label{appendix:saliency}

\RII{Saliency models aim to predict eye fixation points, which are the locations where a person's gaze would typically rest when looking at a scene. Applying saliency models to substitute the usage of a gaze tracker can be an alternative for the VOILA querying paradigm. This section presents the implementation of a saliency-based visual querying system and discusses the experiment results.}

\RII{\subsection{Implementation}}

\RII{Saliency models can be roughly categorized into two types: image and video saliency prediction. For image saliency, we utilized TranSalNet~\cite{TranSalNet}, and for video saliency, we adopted UFO~\cite{su2023unified_UFO}, both leading in the ``Saliency Detection'' task\footnote{https://paperswithcode.com/area/computer-vision}.}

\RII{\begin{itemize}
    \item \textbf{Image Saliency Prediction:} TranSalNet~\cite{TranSalNet} was used to determine saliency areas in selected frames, with the most salient area's center designated as gaze coordinates.
    \item \textbf{Video Saliency Prediction:} UFO's~\cite{su2023unified_UFO} video saliency capabilities allowed for frame-by-frame gaze coordinate prediction, creating a gaze trace analogous to gaze sensor data. This predicted gaze trace was processed using the same pipeline as \system.
\end{itemize}}

\RII{Table~\ref{tab:exp2_result} exclusively presents image-based saliency results, as our system and baselines are implemented with image-level deep learning techniques. As detailed in Appendix~\ref{appendix:g-voila-implement}, transferring entire video clips to \textbf{[Remote Devices]} or executing video-level deep learning on \textbf{[Local Devices]} is impractical. To evaluate video-based saliency models, we pre-transferred all data to our server and conducted preprocessing there.}

\RII{\subsection{Results and Discussion}}

\begin{table}[]
    \centering
    % \resizebox{\linewidth}{!}{
    \begin{tabular}{c|cccccc}
    \toprule
         & \multicolumn{3}{c}{Recall} & \multicolumn{3}{c}{Precision} \\ 
         & All & Explicit & Ambiguous & All & Explicit & Ambiguous \\
    \midrule 
    saliency-image & 78.76\% & 93.86\% & 62.19\% & 72.86\% & 93.03\% & 50.7\% \\
    saliency-video & 78.83\% & 94.12\% & 62.04\% & 73.6\% & 93.31\% & 51.96\% \\
    \bottomrule
    \end{tabular}
    % }
    \caption{\RII{Comparison between VOILA system built on image saliency prediction and video saliency prediction.}}
    \label{tab:saliency_compare}
\end{table}

\RII{Results presented in Tables~\ref{tab:exp2_result} and~\ref{tab:saliency_compare} indicate that the performance of VOILA-center, VOILA-saliency (image), and VOILA-saliency (video) are comparable. VOILA-saliency (video) marginally outperforms VOILA-saliency (image), with the latter showing a slight advantage over VOILA-center, in terms of overall recall and precision. Contrary to expectations, utilizing gaze predictions from saliency models offers a negligible improvement over using head direction (center) to infer interest points. Additionally, temporal saliency analysis (video) does not significantly surpass saliency prediction from isolated frames. We discussed possible reasons, as well as highlighted the gap between current saliency models and the specific demands of \name's querying task.}

\RII{
\begin{itemize}
    \item Why do saliency models underperform compared to using head direction?
    \begin{itemize}
        \item \textbf{Dataset discrepancy:} 
        Saliency models, trained on datasets like SALICON~\cite{jiang2015salicon} and CoCA~\cite{zhang2020coca}, often feature images with eye-catching objects or subjects isolated from the background. In contrast, daily scenarios such as supermarkets present closely arranged homogeneous items, making it hard to identify the user's interested object. Similarly, at-home and street scenarios are composed of complex structure and object placement. Additionally, these models are trained on high-quality photos, whereas our use case involves snapshots from egocentric videos.
        \item \textbf{Semantic gap in gaze for saliency prediction and \name use case:}
        Even though both use cases aim to predict user interest, saliency models emphasize visually striking objects whereas \name indicates objects that users want information for. Therefore, a saliency model might focus on objects with vibrant colors, unique shapes and located in well-illuminated areas, etc. For instance, in the home setting, our employed saliency model tends to highlight a Minions toy over other objects placed on the table. Therefore, as a companion movement to the user's eye movements, the head direction can gain comparable performance with VOILA-saliency. 
    \end{itemize}
    \item Why temporal saliency analysis (video) underperforms frame-level prediction (image):
    \begin{itemize}
        \item The image and video saliency models stem from different research and are not directly comparable.
        \item \textbf{Video saliency prediction requires a more sophisticated temporal locating algorithm because the predicted gaze trace does not share similar properties with real gaze.} Gaze data's feature alone can be used to temporally locate frames that contain the user's interest, as shown in Section~\ref{sec:exp2_objective_score}. However, predicted gaze traces cannot accurately pinpoint interest frames because of both prediction inaccuracies and lower sampling rates. A video saliency model tailored to querying tasks may contribute to this problem. 
        \item \textbf{Static-dynamic mismatch for objects of interest:} Video saliency model typically highlights dynamic objects against a static background~\cite{perazzi2016davis}. However, in scenarios like supermarkets or workspaces, the object of interest is often placed static and captured within a dynamically moving egocentric view.
    \end{itemize}
\end{itemize}}

\end{document}

%\title{\name: Understanding User Behavior with Gaze-Facilitated Information Retrieval in Daily Scenarios}

\title[\name: Gaze-Facilitated Information Querying]{\name: Gaze-Facilitated Information Querying in Daily Scenarios}

% \name: gaze-driven information querying in daily scenarios 


\renewcommand{\shortauthors}{Trovato and Tobin, et al.}


% 整体改成 understanding 类型的文章
% 我们的第一个绿野仙踪实验初步调研了用户在使用 gaze-facilitated QA-based IR 系统时，会展现出怎样的提问方式？更具体的，（1）用户使用的语言将会有怎样的模式变化？（2）用户的语言和眼动行为在时间和空间上将会产生怎样的配合？
% 我们希望能够更进一步地了解用户在真实使用 G-VOILA 时将会产生什么更具体的反思，于是先根据实验一的结果提出了实现 G-VOILA 需要考虑的系统框架，并使用现有的开源技术进行了 naive implementation。
% 讲 evaluation。因为 evaluation 的实验比较复杂，这里需要好好想想怎么能更适配 “understanding” 地把实验讲好。



% \received{15 Nov 2023}

\maketitle

% \begin{acks}
% \end{acks}

\bibliographystyle{ACM-Reference-Format}
\bibliography{sample-base}

\appendix